\documentclass[twocolumn]{jpsj3}
\usepackage{graphicx}
\usepackage{amsmath,amssymb}

\def\runauthor{Youichi {\sc Yanase}}

\title{ 
Random Spin-Orbit Coupling in Spin Triplet Superconductors: 
\\ Stacking Faults in Sr$_2$RuO$_4$ and CePt$_3$Si 
} 

\author{Youichi {\sc Yanase}\footnote{E-mail:
yanase@phys.sc.niigata-u.ac.jp}}

\inst{Department of Physics, Niigata University, Niigata 950-2181, Japan 
}

\recdate{Today 2010}

\abst
{ 
The random spin-orbit coupling in multicomponent superconductors 
is investigated 
focusing on the non-centrosymmetric superconductor CePt$_3$Si and 
the spin triplet superconductor Sr$_2$RuO$_4$. 
 We find novel manifestations of the random spin-orbit coupling 
in the multicomponent superconductors with directional disorders, 
such as stacking faults. 
 The presence of stacking faults is indicated for 
the disordered phase of CePt$_3$Si and Sr$_2$RuO$_4$. 
 It is shown that the $d$-vector of spin triplet 
pairing is locked to be $\vec{d} = k_{\rm y}\hat{x} - k_{\rm x}\hat{y}$ 
with the anisotropy 
$\Delta T_{\rm c}/T_{\rm c0} \sim \bar{\alpha}^{2}/T_{\rm c0}W_{\rm z}$, 
where $\bar{\alpha}$, $T_{\rm c0}$, and $W_{\rm z}$ are the mean square root 
of random spin-orbit coupling, the transition temperature in the clean limit, 
and the kinetic energy along the {\it c}-axis, respectively. 
 This anisotropy is much larger (smaller) than that in the clean bulk 
Sr$_2$RuO$_4$ (CePt$_3$Si). 
 These results indicate that the helical pairing state 
$\vec{d} = k_{\rm y}\hat{x} - k_{\rm x}\hat{y}$ 
in the eutectic crystal of Sr$_2$RuO$_4$-Sr$_3$Ru$_2$O$_7$ is stabilized 
in contrast to the chiral state 
$\vec{d} = (k_{\rm x} \pm {\rm i} k_{\rm y}) \hat{z}$ 
in the bulk Sr$_2$RuO$_4$. 
 The unusual variation of $T_{\rm c}$ in CePt$_3$Si is 
resolved by taking into account the weak pair-breaking effect 
arising from the uniform and random spin-orbit couplings. 
 These superconductors provide a basis for discussing recent topics 
on Majorana fermions and non-Abelian statistics. 
}

\kword
{
spin triplet superconductor, spin-orbit coupling, stacking fault, 
CePt$_3$Si, Sr$_2$RuO$_4$
}

\begin{document}
\sloppy
\maketitle

\newcommand{\eli}{$\acute{{\rm E}}$liashberg }
\renewcommand{\k}{\vec{k}}
\renewcommand{\d}{\vec{d}}
\newcommand{\g}{\vec{g}}
\newcommand{\ktd}{\vec{k}_{\rm 2d}}
\newcommand{\kz}{k_{\rm z}}
\newcommand{\vr}{\vec{r}}
\newcommand{\kk}{\vec{k}'}
\newcommand{\kp}{\vec{k}_{+}}
\newcommand{\kkk}{\vec{k}''}
\newcommand{\q}{\vec{q}}
\newcommand{\Q}{\vec{Q}}
\newcommand{\qp}{\vec{q}_{+}}
\newcommand{\e}{\varepsilon}
\newcommand{\ee}{e}
\newcommand{\s}{{\mit{\it \Sigma}}}
\newcommand{\J}{\mbox{\boldmath$J$}}
\newcommand{\vv}{\mbox{\boldmath$v$}}
\newcommand{\Jh}{J_{{\rm H}}}
\newcommand{\LL}{\mbox{\boldmath$L$}}
\renewcommand{\SS}{\mbox{\boldmath$S$}}
\newcommand{\Tc}{$T_{\rm c}$ }
\newcommand{\Tcf}{$T_{\rm c}$}
\newcommand{\Hc}{$H_{\rm c2}$ }
\newcommand{\Hcf}{$H_{\rm c2}$}
\newcommand{\Pt}{CePt$_3$Si }
\newcommand{\Rh}{CeRhSi$_3$ }
\newcommand{\Ir}{CeIrSi$_3$ }
\newcommand{\Ptf}{CePt$_3$Si}
\newcommand{\Rhf}{CeRhSi$_3$}
\newcommand{\Irf}{CeIrSi$_3$}
\newcommand{\Ru}{Sr$_2$RuO$_4$ }
\newcommand{\Ruf}{Sr$_2$RuO$_4$}
\newcommand{\Rum}{Sr$_3$Ru$_2$O$_7$ }
\newcommand{\Rumf}{Sr$_3$Ru$_2$O$_7$}
\newcommand{\Co}{NaCoO$_2$$\cdot$yH$_2$O }
\newcommand{\Cof}{NaCoO$_2$$\cdot$yH$_2$O}
\newcommand{\Neel}{Ne\'el }

\section{Introduction}

 Spin triplet superconductivity and superfluidity 
have attracted much interest since the discovery of 
multicomponent order parameters in superfluid 
$^{3}$He \cite{Leggett1975} and heavy fermion superconductors.  
\cite{fisher1989,Sigrist1991,Tou1998,Tou1996,Machida1999,Sauls1994,Joynt2002} 
 Recent studies have shown that \Ru \cite{maeno1994}
and non-centrosymmetric \Pt \cite{bauer2004,bauerreview} 
are other candidate spin triplet superconductors. 
 The former is considered to be a $P$-wave superconductor. 
\cite{Mackenzie2003,SigristRu} 
 The mixed-parity $s$+$P$-wave state seems to be realized in the latter 
\cite{sigristncsc,hayashi2006,yanasencsclett,yanasencschelical,yanasencscfull} 
since the crystal structure of \Pt lacks the inversion symmetry. 
\cite{edelstein1989,gorkov-rashba,frigeriprl,fujimotoreview}

 Spin triplet superconductor/superfluid 
has multicomponent order parameters described by the $d$-vector. 
\cite{Leggett1975,Sigrist1991} 
 The structure of the $d$-vector is determined by the spin-orbit coupling 
that breaks the spin SU(2) symmetry. 
 The $d$-vector in the heavy fermion superconductors 
UPt$_{3}$ and UBe$_{13}$ has been investigated on the basis of the 
phenomenological Ginzburg-Landau theory. 
\cite{Sigrist1991,Machida1999,Sauls1994,Joynt2002} 
 Nevertheless, many issues, for instance, the anisotropy of the $d$-vector, 
are still subjects of controversy.

 Triggered by the discovery of superconductivity in \Ruf, 
the microscopic theory of the $d$-vector has been developed. 
 On the basis of the multi-orbital Hubbard model with spin-orbit coupling 
(so-called $L$-$S$ coupling), several microscopic rules for the $d$-vector 
have been obtained, which will be summarized in \S2. \cite{yanaseRu,yanaseCo} 
 According to the microscopic theory, the symmetry-breaking interaction, which 
leads to the anisotropy of the $d$-vector, is very small in many cases. 
 This finding has been confirmed by the nuclear magnetic resonance (NMR) 
measurement of \Ruf. \cite{murakawa2007,murakawa2004}
 The small symmetry-breaking interaction due to the $L$-$S$ coupling 
indicates that another source of spin SU(2) symmetry breaking 
may play an important role in determining the structure of the $d$-vector. 
 The purpose of this study is to investigate the roles of the disorder 
that gives the random spin-orbit coupling. 
 We show that directional disorders such as stacking faults 
significantly affect the $d$-vector in spin triplet superconductors.

 The idea is based on the recent studies on non-centrosymmetric 
superconductors that lack inversion symmetry in their crystal structures. 
\cite{frigeriprl,sigristncsc,hayashi2006,fujimotoreview,bauer2004,bauerreview,yanasencsclett,yanasencschelical,yanasencscfull}  
 It has been shown that antisymmetric spin-orbit coupling plays a 
major role in such systems. \cite{edelstein1989,gorkov-rashba} 
 Although the antisymmetric spin-orbit coupling has the same microscopic 
origin as the $L$-$S$ coupling, \cite{yanasencscfull} 
the effects on the spin triplet superconductivity are considerably different. 
 The effect of the antisymmetric spin-orbit coupling 
is much larger than that of the $L$-$S$ coupling, 
since the splitting of Fermi surfaces is induced by the former. 
 Thus, we are led to the idea that the antisymmetric spin-orbit coupling 
may also play an important role in the {\it globally} centrosymmetric 
system with a broken {\it local} inversion symmetry. 
 We show that this is the case for a spin triplet superconductor 
in the presence of directional disorders.

 The presence of stacking faults in the eutectic crystal of 
\Ruf-\Rum \cite{kittaka2008,kittaka2009,fittipaldi2008} 
as well as in the disordered phase of \Pt has been pointed out. 
\cite{mukuda2009} 
 The former is regarded to be a disordered phase of the centrosymmetric 
superconductor \Ruf, while the latter is a disordered phase of 
the non-centrosymmetric superconductor. 
\cite{bauerreview,takeuchi2007,motoyama2008} 
 The local inversion symmetry is broken in these materials, while the 
global inversion symmetry is recovered by the randomness. 
 We investigate the superconductivity in these systems by assuming 
the random spin-orbit coupling and random scalar potential 
arising from stacking faults. 

 We show that the $d$-vector in \Ruf-\Rum is different from 
the chiral state in the bulk \Ruf. \cite{Mackenzie2003} 
 The eutectic \Ruf-\Rum is a time-reversal invariant spin-triplet 
superconductor that attracts much attention in terms of 
Majorana fermions, non-Abelian statistics and their relationship 
with topological properties. 
\cite{read-green,ivanov2001,schnyder2008,roy2008,qi2009,sato2009}

 In this study, we also resolve the seemingly controversial 
issue of the non-centrosymmetric superconductor \Ptf. 
 The \Tc of the disordered \Pt is higher than that 
of the clean \Ptf. \cite{mukuda2009,takeuchi2007,motoyama2008,bauerreview} 
 This variation of \Tc is incompatible with the usual pair-breaking effect 
in non-$s$-wave superconductors. 
 We show that this unusual variation of \Tc is attributed to 
the pair-breaking effect arising from the antisymmetric spin-orbit coupling.

 The paper is organized as follows. In \S2, we summarize the results 
obtained using the microscopic theory for the $d$-vector in clean spin triplet 
superconductors. 
 The effects of the {\it uniform} spin-orbit coupling are discussed 
for both centrosymmetric and non-centrosymmetric systems. 
 In \S3, we formulate the {\it random} spin-orbit coupling arising from 
stacking faults. 
 The eutectic \Ruf-\Rum and the disordered \Pt are modeled 
in a unified way.  
 The effects of the random spin-orbit coupling as well as the random scalar 
potential are investigated on the basis of the self-consistent 
Born approximation. 
 The results for the $d$-vector and the pair-breaking effects 
are shown in \S4. 
 The breakdown of the Born approximation in the highly two-dimensional system  
is pointed out in \S4.3, where the results expected in the two-dimensional 
limit are shown. 
 The superconductivities in \Pt and \Ru are discussed in \S5.1 and \S5.2, 
respectively. 
 The $d$-vectors in the spin triplet superconductors are summarized in \S6. 
A discussion is given in \S7.

\section{$D$-vector in Clean Spin Triplet Superconductors}

 The discovery of superconductivity in \Ru \cite{maeno1994} 
led to a breakthrough in the microscopic theory of 
spin triplet superconductivity, 
since the simple electronic structure of \Ru made it possible to study 
the $d$-vector on the basis of microscopic models, 
such as the multi-orbital Hubbard model \cite{yanaseRu,nomuraprivate} 
and multi-orbital $d$-$p$ model \cite{yoshioka2009,nomuraprivate}.

 One of the achievements of the microscopic theory is 
the formulation of rules for the $d$-vector in $d$-electron systems such as 
\Ru \cite{yanaseRu} and \Cof, \cite{yanaseCo} 
which are summarized in Table I. 
 Since these superconductors have inversion symmetry, the 
spin-orbit coupling is described by the so-called $L$-$S$ coupling $\lambda$.  
 The relation $T_{\rm c} \ll \lambda \ll E_{\rm F}$ holds in the $d$-electron 
systems with $E_{\rm F}$ being the Fermi energy.  
 Since the large parameter $\lambda/T_{\rm c}$ is irrelevant in the presence 
of inversion symmetry, the perturbative treatment 
with respect to the small parameter $\lambda/E_{\rm F}$ 
is justified. \cite{yanaseRu} 
 On the basis of this fact, we classified the structures of the $d$-vector 
shown in Table I. 

 We found that the $d$-vector is determined in many cases solely by the 
symmetries of the crystal structure (first row), local electron orbital 
(second row), and superconductivity (third row). 
 The direction and anisotropy of the $d$-vector are shown 
in the fourth and fifth rows, respectively. 
 Here, the anisotropy is defined as 
$\Delta T_{\rm c}/T_{\rm c} = (T_{\rm c}-T_{\rm c}^{*})/T_{\rm c}$ 
with $T_{\rm c}$ and $T_{\rm c}^{*}$ being the transition temperatures 
for the most and second most stable pairing states, respectively. 
 Most of these results are exact in the sense that they are 
independent of the details of the Fermi surface and electron correlation. 
 This is because the selection rules due to the symmetries solely determine 
the effect of spin-orbit coupling in the lowest order of $O(\lambda/E_{\rm F})$. 
\cite{yanaseCo} 

 As an exceptional case, the direction of the $d$-vector is not exactly 
determined when the lowest order term of $\lambda/E_{\rm F}$ vanishes. 
 In such a case, the anisotropy is as small as $O(\lambda^{2}/E_{\rm F}^{2})$. 
 This is the case for \Ru, where the superconductivity is mainly induced by 
the d$_{\rm xy}$-orbital electrons. 
 We determined the $d$-vector for the d$_{\rm xy}$-orbital electrons 
in the tetragonal lattice on the basis of the perturbation theory for 
the three-orbital Hubbard model. \cite{yanaseRu} 
 Then, we found that the $d$-vector indeed depends on microscopic details 
such as the Fermi surface and electron correlation. 
 We assume the band structure of \Ru obtained by the band calculation 
and show our result in the corresponding part of Table I.

 We here give two additional comments on Table I. 
 First, the symmetries of the crystal structure and superconductivity 
are taken into account in the phenomenological Ginzburg-Landau theory, 
\cite{Sigrist1991} while the microscopic theory is needed to take 
advantage of the symmetry of the local electron orbital. 
 In other words, the local orbital plays an essential role in obtaining the 
results summarized in Table I.

\begin{table}
\begin{tabular}[htb]{|c|c|c|c|c|} \hline
\multicolumn{2}{|c|}{Tetragonal} 
& \multicolumn{3}{|c|}{Hexagonal} 
\\ \hline  
d$_{\rm xy}$ & d$_{\rm xz}$, d$_{\rm yz}$ & 
\multicolumn{2}{|c|}{E$_{\rm g}$} & A$_{\rm 1g}$ 
\\ \hline
\multicolumn{2}{|c|}{P-wave} & P-wave & F-wave & P- or F-wave
\\ \hline \hline 
$ \d \parallel c$ & $ \d \parallel ab$ & $ \d \parallel ab$ & both 
& both 
\\ \hline
$O(\lambda^{2}/E_{\rm F}^{2})$ & $O(\lambda/E_{\rm F})$ & 
$O(\lambda/E_{\rm F})$ & $O(\lambda^{2}/E_{\rm F}^{2})$ & $O(\lambda^{2}/E_{\rm F}^{2})$
\\ \hline
\end{tabular}
\caption{Summary of the $d$-vector in the clean centrosymmetric spin triplet 
superconductors. \cite{yanaseRu,yanaseCo} 
The direction (fourth row) and anisotropy (fifth row) of the $d$-vector 
are determined by the symmetries of the crystal structure (first row), 
local electron orbital (second row), and superconductivity (third row). 
See the text for details. 
}
\end{table}

 Second and more importantly, the anisotropy of the $d$-vector is generally 
small when the spin-orbit coupling is smaller than the Fermi energy 
$\lambda < E_{\rm F}$. \cite{yanaseRu} 
 This is true even when the spin-orbit coupling is much larger 
than \Tcf. 
 This means that the $d$-vector is rotated by a small applied magnetic field 
parallel to the $d$-vector, as confirmed by NMR measurements of \Ruf. 
\cite{murakawa2004,murakawa2007}

 Another category of spin triplet superconductors is the 
non-centrosymmetric superconductors. 
 The order parameter of such superconductors cannot 
be classified into even parity or odd parity 
because of the broken inversion symmetry. 
 Then, an admixture of spin singlet and spin triplet Cooper pairs 
occurs, and therefore the order parameter of spin triplet pairing is 
always finite. \cite{edelstein1989}  
 The theory of the $d$-vector for such systems is rather simple since the 
splitting of Fermi surfaces is induced by the antisymmetric spin-orbit 
coupling. 
 It has been shown that the $d$-vector is parallel to the $g$-vector 
that characterizes the antisymmetric spin-orbit coupling. \cite{frigeriprl} 
 Since the symmetry of the $g$-vector is determined by the 
crystal structure, the $d$-vector is determined solely by the 
crystal symmetry. 
 In the case of the P$_{4mm}$ space group of \Ptf, the $g$-vector 
is of the Rashba-type \cite{rashba1959} 
and then, the $d$-vector is $\d(\k) = k_{\rm y}\hat{x} - k_{\rm x}\hat{y}$.

 The coupling constant of the antisymmetric spin-orbit coupling $\alpha$ 
satisfies the relation $T_{\rm c} \ll \alpha \ll E_{\rm F}$ in most 
non-centrosymmetric superconductors, including heavy fermion systems
such as \Ptf, \Rhf, and \Irf. 
 This relation is similar to that for the $L$-$S$ coupling $\lambda$ 
for the centrosymmetric $d$-electron systems. 
 However, the antisymmetric spin-orbit coupling gives rise to much larger 
anisotropy of the $d$-vector than the $L$-$S$ coupling, 
because the large parameter $\alpha/T_{\rm c}$ is relevant for the 
superconductivity. 
 This means that the antisymmetric spin-orbit coupling plays a major role 
in the spin triplet superconductivity even when its coupling constant is 
much smaller than the $L$-$S$ coupling.  
 This is the reason why we focus on the 
{\it random antisymmetric spin-orbit coupling} in the disordered system 
and ignore the $L$-$S$ coupling in this paper.

\section{Random Spin-Orbit Coupling in Stacking Fault Model}

\subsection{Stacking fault model}

 We first formulate the random spin-orbit coupling and random scalar potential 
arising from stacking faults. Assuming the stacking fault model for 
disordered \Pt \cite{mukuda2009} 
and eutectic \Ruf-\Rumf, \cite{kittaka2008,kittaka2009,fittipaldi2008} 
the crystal structures of these materials are schematically shown in Fig.~1. 
 \Pt lacks the inversion symmetry in the clean limit, 
but the global inversion symmetry is restored by stacking faults 
while keeping the broken local inversion symmetry [Fig.~1(a)]. 
 On the other hand, clean \Ru has the inversion symmetry, 
while stacking faults lead to local inversion symmetry breaking 
while keeping the global inversion symmetry [Fig.~1(b)]. 
 Thus, \Pt and \Ru are contrasting examples 
and can be investigated in a unified way.

\begin{figure}[htbp]
  \begin{center}
\includegraphics[width=7cm]{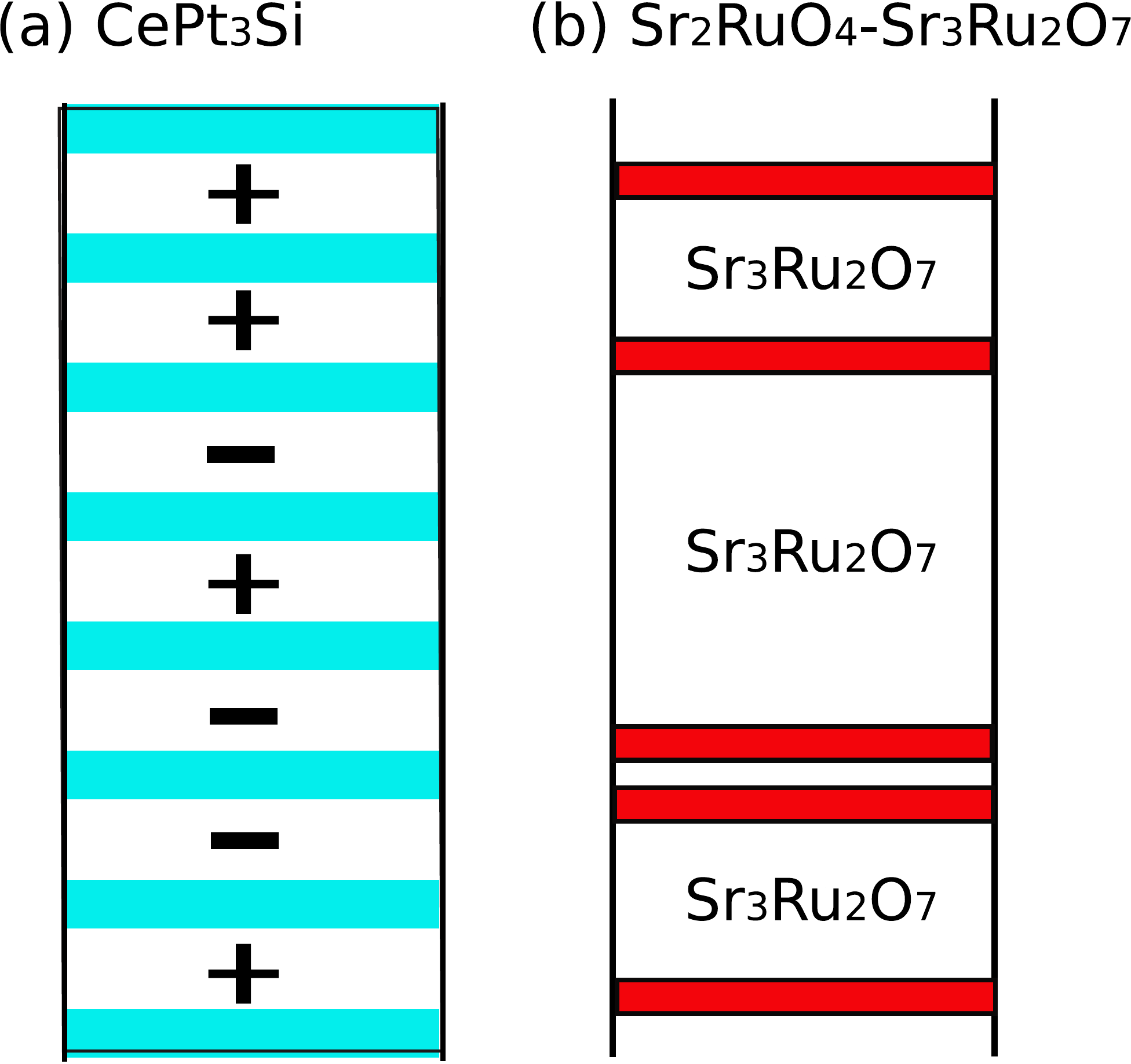}
\caption{(Color online) 
Schematic figures of the stacking faults in (a) \Pt and (b) \Ruf-\Rumf.
(a) Filled boxes show the layers of Ce atoms. The Si and Pt atoms fill the 
space between the layers. The two possible positions of Si atoms are 
described by $+$ and $-$. The randomness is induced by the random distribution 
of $+$ and $-$ blocks. 
(b) Filled boxes show the RuO$_2$ layers in which the superconductivity occurs. 
Most of the spatial region is filled by the metamagnet \Rumf. 
The randomness arises from the random distribution of RuO$_2$ layers. 
}
    \label{fig:stackingfault}
  \end{center}
\end{figure}

 To focus on stacking faults in the spin triplet superconductors, 
we assume a three-dimensional model in which each two-dimensional layer 
is clean but the stacking along the {\it c}-axis is disordered 
as in Fig.~1. 
 The model is described as 
\begin{eqnarray}
\label{eq:model}
&& \hspace*{-7mm}  H = \sum_{\k,s} \e(\k) c_{\k,s}^{\dag}c_{\k,s} 
   + \sum_{\gamma=1}^{6} g_{\gamma} \lambda_{\gamma}^{\dag} \lambda_{\gamma} 
   + \sum_{\vr, i} u_{i} n_{\vr,i} 
\nonumber \\
&& \hspace*{-0mm}   
   + \sum_{\ktd,i} \alpha_{i} \vec{g}(\ktd) \cdot \vec{S}_{i}(\ktd),   
\end{eqnarray}
where $\k = (k_{\rm x},k_{\rm y},k_{\rm z})$ and $\ktd = (k_{\rm x},k_{\rm y})$ 
represent the three- and two-dimensional momenta, respectively. 
We denote the index of each layer as $i$ and the spin as $s$. We denote 
$\vec{S}_{i}(\ktd) = \sum_{s,s'} \vec{\sigma}_{ss'} c_{\ktd,i,s}^{\dag}c_{\ktd,i,s'}$, 
with 
$\vec{\sigma}$ being the vector representation of the Pauli matrix.   
$n_{\vr,i}$ is the electron number at the site $(\vr,i)$. 
 The creation operators of spin triplet Cooper pairs are described as 
$\lambda_{\gamma}^{\dag} = \sum_{\k,s,s'} 
\d_{\gamma}(\k)({\rm i}\vec{\sigma}\sigma_{\rm y})_{ss'} c_{\k,s}^{\dag}c_{-\k,s'}^{\dag}$. 
where $\d_{1,2}(\k) = \frac{1}{\sqrt{2}} (\phi_{\rm x}(\k),\pm \phi_{\rm y}(\k),0)$,
      $\d_{3,4}(\k) = \frac{1}{\sqrt{2}} (\phi_{\rm y}(\k),\pm \phi_{\rm x}(\k),0)$,
and   $\d_{5,6}(\k) = \frac{1}{\sqrt{2}} 
       (0,0,\phi_{\rm x}(\k) \pm {\rm i} \phi_{\rm y}(\k))$ are the irreducible 
representations of the order parameter 
in the tetragonal lattice. \cite{Sigrist1991} 
 We denote the $d$-vectors of these states 
$\d = k_{\rm x} \hat{x} \pm k_{\rm y} \hat{y}$, 
$\d = k_{\rm y} \hat{x} \pm k_{\rm x} \hat{y}$, 
and $\d = (k_{\rm x} \pm {\rm i} k_{\rm y}) \hat{z}$,  
respectively. 
 Although the effect of $L$-$S$ coupling is included in the 
differences of pairing interactions $g_{\gamma}$ for each pairing state 
($\gamma = 1-6$), we here assume $g_{\gamma} = g$ for simplicity. 
 This means that the effect of $L$-$S$ coupling is ignored. 
 This simplification is valid when the effect of $L$-$S$ coupling is small 
as mentioned earlier.

 The random scalar potential and random spin-orbit coupling at layer 
$i$ are represented by $u_{i}$ and $\alpha_{i}$ in the third and fourth terms 
in eq.~(\ref{eq:model}), respectively. 
 The random variables $u_{i}$ and $\alpha_{i}$ are independent of 
the two-dimensional coordinate $\vr$ in the stacking fault model. 
 The random spin-orbit coupling $\alpha_{i}$ arises from the local mirror  
symmetry breaking with respect to the two-dimensional plane. 
 We assume the random averages, $<u_{i}> = <\alpha_{i}> = 0$, 
$<u_{i}u_{j}> = \bar{u}^{2} \delta_{i,j}$, 
$<\alpha_{i}\alpha_{j}> = \bar{\alpha}^{2} \delta_{i,j}$, 
and $<u_{i}\alpha_{j}> = 0$.

\subsection{Born approximation}

\begin{figure}[htbp]
  \begin{center}
\includegraphics[width=6cm]{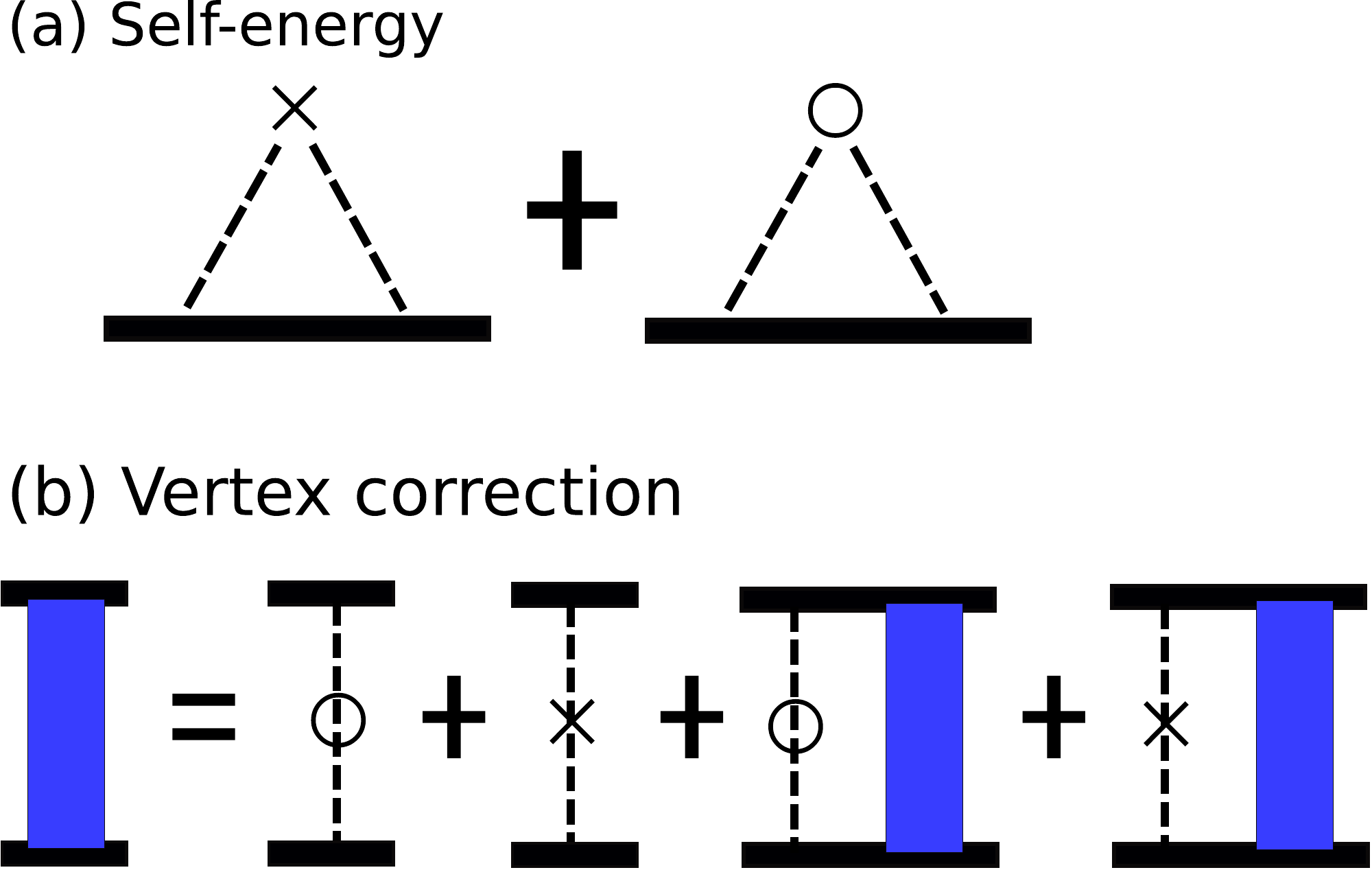}
\caption{(Color online) 
 Diagrammatic representations of (a) self-energy and (b) vertex correction 
to the irreducible susceptibility in the Born approximation. 
 The crosses and circles represent the scattering due to the 
random scalar potential and random spin-orbit coupling, respectively. 
}
    \label{fig:diagrams}
  \end{center}
\end{figure}

 We solve the model given by eq.~(1) on the basis of the Born approximation 
by assuming $\bar{u}, \bar{\alpha} \ll W_{\rm z}$, with $W_{\rm z}$ 
being the kinetic energy along the {\it c}-axis. 
 The diagrammatic representations of the self-energy and the vertex correction 
to the irreducible susceptibility are shown in Figs.~2(a) and 2(b), respectively. 

 The Green function and self-energy are obtained as 
\begin{eqnarray}
\label{eq:Green-function}
&& \hspace*{-7mm}  
G(\k,\omega_{n})^{-1} = G^{0}(\k,\omega_{n})^{-1} - \Sigma(\ktd,\omega_{n}), 
\\ && \hspace*{-7mm}  
\Sigma(\ktd,\omega_{n}) = (\bar{u}^{2} + \bar{\alpha}^{2} |\g(\ktd)|^{2})
\sum_{\kz} G(\k,\omega_{n}), 
\end{eqnarray}
respectively. 
 The undressed Green function is given as 
$G^{0}(\k,\omega_{n}) = ({\rm i}\omega_{n} - \e(\k))^{-1}$, where 
$\omega_{n} = (2 n +1) \pi T$ is the Matsubara frequency and $T$ is the 
temperature.

 The irreducible susceptibility is divided into the contributions from 
the intralayer and interlayer pairings, 
\begin{eqnarray}
\label{eq:susceptibility}
&& \hspace*{-7mm} 
\chi_{\rm sc} = T \sum_{\ktd,\omega_{n}} 
[T^{\rm 2d}(\ktd,\omega_n) + T^{\rm 3d}(\ktd,\omega_n)], 
\end{eqnarray}
where 
\begin{eqnarray}
\label{eq:T-matrix}
&& \hspace*{-10mm}  
T^{\rm 2d}(\ktd,\omega_n) = 
|\d_{\rm 2d}(\ktd,\omega_n)|^{2} 
\{|\hat{d}_{\rm 2d}(\ktd,\omega_n) \cdot \hat{g}(\ktd)|^{2}
\nonumber \\ && \hspace*{8mm}  \times
1/[t(\ktd,\omega_n)^{-1} - (\bar{u}^{2} + \bar{\alpha}^{2} |\g(\ktd)|^{2})]
\nonumber \\ && \hspace*{8mm}  
+(1-|\hat{d}_{\rm 2d}(\ktd,\omega_n) \cdot \hat{g}(\ktd)|^{2})
\nonumber \\ && \hspace*{8mm}  \times
1/[t(\ktd,\omega_n)^{-1} - (\bar{u}^{2} - \bar{\alpha}^{2} |\g(\ktd)|^{2})]
\}, 
\\ && \hspace*{-10mm}  
T^{\rm 3d}(\ktd,\omega_n) = \sum_{\kz} 
|\d_{\rm 3d}(\k,\omega_n)|^{2} 
|G(\k,\omega_n)|^{2},  
\end{eqnarray} 
and $t(\ktd,\omega_n) = \sum_{\kz} |G(\k,\omega_n)|^{2}$. 
 We denote the unit vectors as $\hat{a} = \vec{a}/|\vec{a}|$. 
 We separate the $d$-vector into the wave functions of the intralayer 
and interlayer Cooper pairings as 
$\d(\k) = \d_{\rm 2d}(\ktd,\omega_n) + \d_{\rm 3d}(\k,\omega_n)$ so as to 
satisfy the following relations; 
\begin{eqnarray}
\label{eq:phi_3d}
&& \hspace*{-10mm}  
\d_{\rm 2d}(\ktd,\omega_n) \sum_{\kz}  |G^{0}(\k,\omega_{n})|^{2} 
= 
\sum_{\kz} \d(\k) |G^{0}(\k,\omega_{n})|^{2}, 
\\ && \hspace*{-10mm}
\label{eq:phi_2d}
\sum_{\kz} \d_{\rm 3d}(\k,\omega_n) |G^{0}(\k,\omega_{n})|^{2} = 0. 
\end{eqnarray}
 When the $d$-vector is $\kz$-independent, the three-dimensional component 
of the $d$-vector vanishes as $\d_{\rm 3d}(\k,\omega_n)=0$. 
 When the $d$-vector is $\kz$-dependent and even with respect to $\kz$, 
namely, $\d(\ktd,\kz) = \d(\ktd,-\kz)$, 
the three-dimensional component of the $d$-vector $\d_{\rm 3d}(\k,\omega_n)$ 
changes its sign at a finite $\kz$. 
 When the $d$-vector is odd with respect to $\kz$, 
the two-dimensional component of the $d$-vector vanishes as 
$\d_{\rm 2d}(\ktd,\omega_n) =0$, and therefore $\d_{\rm 3d}(\k,\omega_n) = \d(\k)$.
 We focus on the $d$-vector having even $\kz$ dependence in the following 
part, and give a brief comment on the odd $d$-vector with respect to $\kz$ 
in \S4.1 and \S4.2.

 \Tc is determined by the criterion 
\begin{eqnarray}
\label{eq:Tc}
&& \hspace*{-7mm} 
\chi_{\rm sc} (T=T_{\rm c},\bar{u},\bar{\alpha}) = 
\chi_{\rm sc} (T=T_{\rm c0},\bar{u}=0,\bar{\alpha}=0), 
\end{eqnarray}
where $T_{\rm c0}$ is the transition temperature in the clean limit 
($\bar{u}=0$ and $\bar{\alpha}=0$).

\subsection{Weak-coupling theory}

 In this subsection, we solve eqs.~(\ref{eq:Green-function})-(\ref{eq:Tc}) 
on the basis of weak-coupling theory. 
Assuming $(\bar{u}^{2}+\bar{\alpha}^{2})/W_{\rm z} \ll E_{\rm F}$,  
the self-energy is obtained as 
\begin{eqnarray}
\label{eq:ap-Self-energy}
&& \hspace*{-7mm} 
\Sigma(\ktd,\omega_{n}) = -{\rm i} (\Gamma^{u}(\ktd) + \Gamma^{\alpha}(\ktd)), 
\\ && \hspace*{-7mm}  
\Gamma^{u}(\ktd) = \pi \bar{u}^{2} \rho^{\rm z} (\ktd),
\\ && \hspace*{-7mm}  
\Gamma^{\alpha}(\ktd) = \pi \bar{\alpha}^{2} |\g(\ktd)|^{2} \rho^{\rm z} (\ktd),
\end{eqnarray}
where $\rho^{\rm z} (\ktd) = \sum_{\kz} \delta(\e(\k))$. 
 Using $G(\k,\omega_{n})^{-1}={\rm i} \bar{\omega}_{n}(\ktd) - \e(\k)$ with 
$\bar{\omega}_{n}(\ktd) = \omega_{n} + \Gamma^{u}(\ktd) + \Gamma^{\alpha}(\ktd)$, 
we obtain 
$t(\ktd,\omega_n) = \pi \rho^{\rm z} (\ktd)/\bar{\omega}_{n}(\ktd)$, 
and therefore, 
\begin{eqnarray}
\label{eq:ap-T-matrix}
&& \hspace*{-10mm} 
T^{\rm 2d}(\ktd,\omega_n) = |\d_{\rm 2d}(\ktd)|^{2} 
[|\hat{d}_{\rm 2d}(\ktd) \cdot \hat{g}(\ktd)|^{2} 
\pi \rho^{\rm z} (\ktd)/\omega_{n} 
\nonumber \\ && \hspace*{-0mm}  
+ (1 - |\hat{d}_{\rm 2d}(\ktd) \cdot \hat{g}(\ktd)|^{2}) 
\pi \rho^{\rm z} (\ktd)/\bar{\omega}_{n}^{1}(\ktd)], 
\\ && \hspace*{-10mm}  
T^{\rm 3d}(\ktd,\omega_n) = \sum_{\kz} |\d_{\rm 3d}(\k)|^{2}
\pi \delta(\e(\k))/\bar{\omega}_{n}(\ktd), 
\end{eqnarray}
where 
$\d_{\rm 2d, 3d}(\ktd)=\d_{\rm 2d, 3d}(\ktd,\omega_n)|_{\omega_n \rightarrow 0}$ 
and  
$\bar{\omega}_{n}^{1}(\ktd) = \omega_{n} + 2 \Gamma^{\alpha}(\ktd)$. 
 Following eq.~(4), we obtain the irreducible susceptibility 
for the superconductivity, 
\begin{eqnarray}
\label{eq:ap-susceptibility}
&& \hspace*{-12mm} 
\chi_{\rm sc} - \chi_{\rm sc}(\bar{u}=0,\bar{\alpha}=0) = 
\sum_{\k} |\d_{\rm 3d}(\k)|^{2} \delta(\e(\k)) 
\nonumber \\ && \hspace*{5mm} \times
[\psi(\frac{1}{2}) - \psi(\frac{1}{2} 
+ \frac{\Gamma^{u}(\ktd)+\Gamma^{\alpha}(\ktd)}{2 \pi T})]
\nonumber \\ && \hspace*{-5mm} 
+ \sum_{\ktd} |\d_{\rm 2d}(\ktd)|^{2} 
(1 - |\hat{d}_{\rm 2d}(\ktd) \cdot \hat{g}(\ktd)|^{2}) 
\rho^{\rm z} (\ktd)
\nonumber \\ && \hspace*{15mm} \times
[\psi(\frac{1}{2}) - \psi(\frac{1}{2}+\frac{2 \Gamma^{\alpha}(\ktd)}{2 \pi T})], 
\end{eqnarray}
where 
$\psi(x)$ is the digamma function.

 According to eq.~(9), the transition temperature is determined as 
\begin{eqnarray}
\label{eq:ap-Tc}
&& \hspace*{-10mm} 
\rho_{\rm t} \log\frac{T_{\rm c}}{T_{\rm c0}} = 
\nonumber \\ && \hspace*{-10mm} 
\sum_{\k} |\d_{\rm 3d}(\k)|^{2} \delta(\e(\k)) 
[\psi(\frac{1}{2}) - \psi(\frac{1}{2} 
+ \frac{\Gamma^{u}(\ktd)+\Gamma^{\alpha}(\ktd)}{2 \pi T_{\rm c}})]
\nonumber \\ && \hspace*{-10mm} 
+ \sum_{\ktd} |\d_{\rm 2d}(\ktd)|^{2} 
(1 - |\hat{d}_{\rm 2d}(\ktd) \cdot \hat{g}(\ktd)|^{2}) \rho^{\rm z}(\ktd)
\nonumber \\ && \hspace*{20mm} \times
[\psi(\frac{1}{2}) 
- \psi(\frac{1}{2}+\frac{2 \Gamma^{\alpha}(\ktd)}{2 \pi T_{\rm c}})], 
\end{eqnarray}
where $\rho_{\rm t} = \sum_{\k} |\d(\k)|^{2} \delta(\e(\k))$.

 Using a similar analysis, we obtain the equation of \Tc 
for the spin singlet pairing state as 
\begin{eqnarray}
\label{eq:ap-Tc-singlet}
&& \hspace*{-8mm} 
\rho_{\rm s} \log\frac{T_{\rm c}}{T_{\rm c0}} = 
\nonumber \\ && \hspace*{-8mm} 
\sum_{\k} |\phi^{\rm s}_{\rm 3d}(\k)|^{2} \delta(\e(\k)) 
[\psi(\frac{1}{2}) - \psi(\frac{1}{2} 
+ \frac{\Gamma^{u}(\ktd)+\Gamma^{\alpha}(\ktd)}{2 \pi T_{\rm c}})], 
\nonumber \\ &&
\end{eqnarray}
where $\rho_{\rm s} = 
\sum_{\k} |\phi^{\rm s}(\k)|^{2} \delta(\e(\k))$. 
 The scalar order parameter of the spin singlet superconductivity 
is denoted as $\phi^{\rm s}(\k) = \phi^{\rm s}(-\k)$ and its interlayer component 
$\phi^{\rm s}_{\rm 3d}(\k)$ is defined in the same way as 
in eqs.~(\ref{eq:phi_3d}) and (\ref{eq:phi_2d}).

\section{$D$-vector and Pair-Breaking Effect}

 In this section, we investigate the effects of random spin-orbit coupling 
and random scalar potential on the spin triplet superconductors. 
 The effects on the $d$-vector are clarified in \S4.1 and 
the pair-breaking effect is investigated in \S4.2.

\subsection{$D$-vector}

 First, we discuss the $d$-vector in the presence of 
random spin-orbit coupling. 
 The pair-breaking effect arising from the random spin-orbit coupling 
leads to the anisotropy of the $d$-vector 
through the second term on the right-hand side of eq.~(\ref{eq:ap-Tc}). 
 We see that the anisotropy originates from the intralayer Cooper pairing 
represented by $\d_{\rm 2d}(\ktd)$. 
 The interlayer Cooper pairs are suppressed by disorders 
independent of the spin degree of freedom. 
 Since the scalar disorder does not give rise to the anisotropy of the 
$d$-vector, we focus on the random spin-orbit coupling in this subsection.

 Because the second term in eq.~(\ref{eq:ap-Tc}) vanishes 
for the $d$-vector parallel to the $g$-vector,   
the spin triplet superconducting state with $\d(\k) \parallel \g(\ktd)$ 
is robust against the random spin-orbit coupling. 
 The other spin triplet pairing states are destabilized 
by the random spin-orbit coupling. 
 The $d$-vector $\d = k_{\rm y} \hat{x} - k_{\rm x} \hat{y}$ is favored 
when the random spin-orbit coupling is of the Rashba type. 
 This is the same pairing state as that in clean 
non-centrosymmetric superconductors having the spatially uniform 
Rashba spin-orbit coupling. \cite{frigeriprl} 
 On the other hand, the anisotropy of the $d$-vector is significantly 
different between the clean and disordered systems. 
 The pair-breaking effect on the spin triplet pairing state with 
$\d(\k) \not\parallel \g(\ktd)$ is represented by the parameter 
$\Gamma^{\alpha}/T_{\rm c0}$. 
 The phase relaxation rate $\Gamma^{\alpha}$ is obtained as 
$\Gamma^{\alpha} \sim \alpha$ in the clean non-centrosymmetric 
superconductors, \cite{frigeriprl} 
while $\Gamma^{\alpha} \sim \pi \bar{\alpha}^{2}/W_{\rm z}$ in the 
stacking fault model. 
 The anisotropy arising from the antisymmetric 
spin-orbit coupling is significantly decreased by stacking faults 
when the relation $\bar{\alpha} \ll W_{\rm z}$ is satisfied, as in 
most non-centrosymmetric superconductors. 
 This is one of the manifestations of the global inversion symmetry 
recovered by the disorders.

\begin{figure}[htbp]
  \begin{center}
\includegraphics[width=7cm]{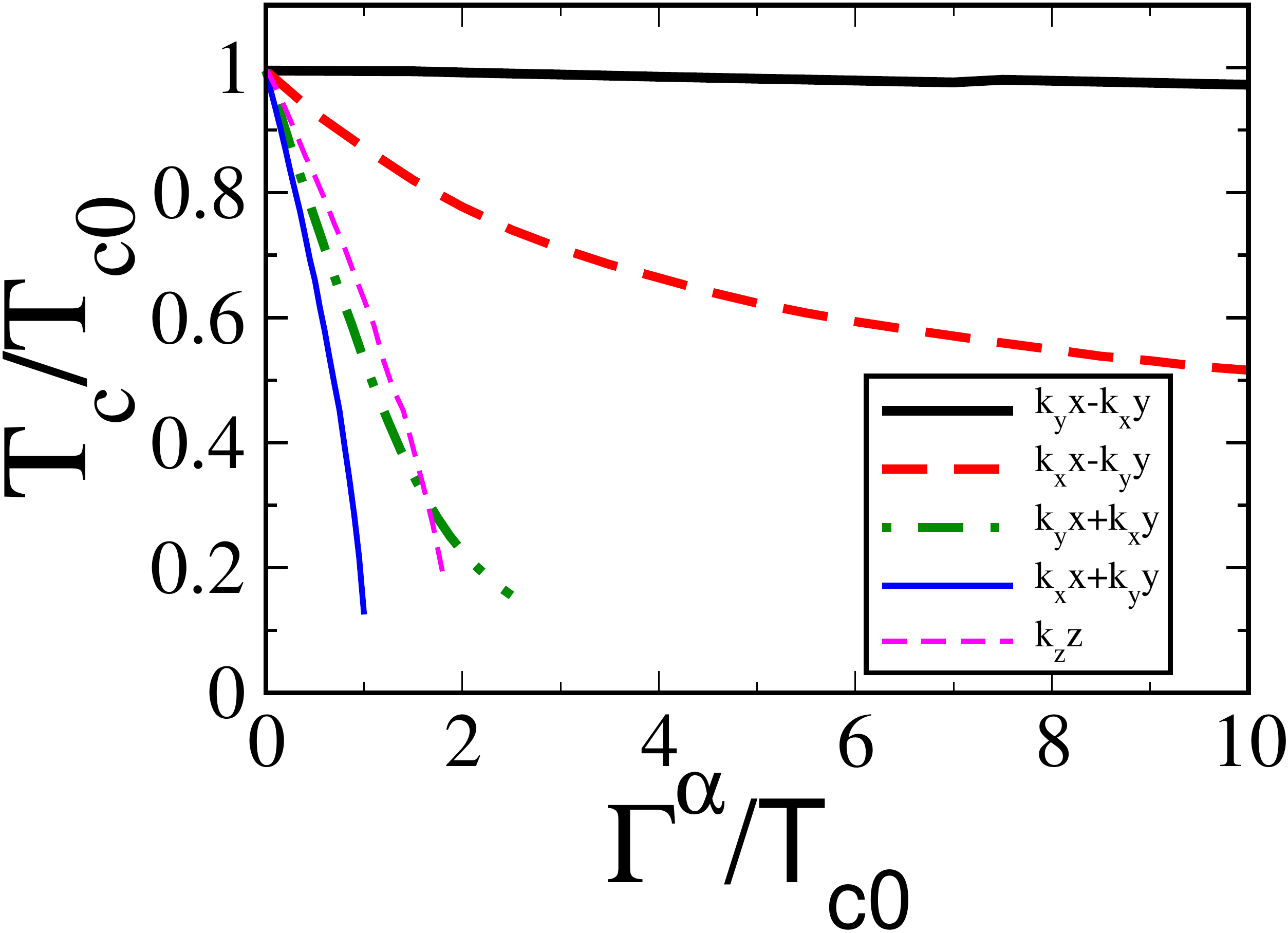}
\caption{(Color online) 
\Tc values of various spin triplet pairing states in the presence of
random spin-orbit coupling. 
We plot the normalized quantity $T_{\rm c}/T_{\rm c0}$ 
for the dimensionless parameter 
$\Gamma^{\alpha}/T_{\rm c0} = \bar{\alpha}^{2}/T_{\rm c0}t_{\rm z}$. 
We assume $T_{\rm c0}=0.0256$. 
The thick solid, thick dashed, dash-dotted, and thin solid 
lines show the \Tc values of the pairing states 
$\d = k_{\rm y} \hat{x} - k_{\rm x} \hat{y}$, 
$\d = k_{\rm x} \hat{x} - k_{\rm y} \hat{y}$, 
$\d = k_{\rm y} \hat{x} + k_{\rm x} \hat{y}$, 
and $\d = k_{\rm x} \hat{x} + k_{\rm y} \hat{y}$, 
respectively. 
The \Tc of the chiral state 
$\d = (k_{\rm x} \pm {\rm i} k_{\rm y}) \hat{z}$ is the same as 
that of $\d = k_{\rm x} \hat{x} + k_{\rm y} \hat{y}$. 
The scalar disorder with 
$\Gamma^{u}/T_{\rm c0} = \bar{u}^{2}/T_{\rm c0}t_{\rm z} = 5$ 
is taken into account, but its effect on \Tc is negligible. 
 We also show the \Tc of the pairing state $\d = k_{\rm z} \hat{z}$ for 
$\Gamma^{u}/T_{\rm c0} = 0$ (thin dashed line).  
}
  \end{center}
\end{figure}

 Figure~3 shows the \Tc values of various spin triplet pairing states 
for the simple dispersion relation  
\begin{eqnarray}
\label{eq:dispersion}
&& \hspace*{-8mm}  \e(\k)  =   2 t (\cos k_{\rm x} +\cos k_{\rm y}) 
 + 2 t_{\rm z} \cos k_{\rm z} - \mu, 
\end{eqnarray}
and the $g$-vector 
$\g(\k) = (-\sin k_{\rm y}, \sin k_{\rm x}, 0)/<|\g(\k)|>_{\rm F}$, 
where the bracket $<>_{\rm F}$ means the average on the Fermi surface. 
 We numerically solve eqs.~(\ref{eq:Green-function})-(\ref{eq:Tc}) 
without using the weak-coupling approximation in \S3.3, 
although we have confirmed that the weak-coupling approximation leading 
to eq.~(\ref{eq:ap-Tc}) is quantitatively valid. 
 We choose the unit of energy as $t=1$ and assume $t_{\rm z}=0.2$. 
 The chemical potential $\mu$ is determined so that the electron density 
per site is $n=0.5$. 
 The wave functions of Cooper pairs are assumed to be 
$(\phi_{\rm x}(\k),\phi_{\rm y}(\k)) = (\sin k_{\rm x},\sin k_{\rm y})$. 
 For this model, $\d_{\rm 3d}(\k) = 0$, and therefore, 
the first term of eq.~(\ref{eq:ap-Tc}) vanishes.

 We see that the spin triplet pairing state 
$\d = k_{\rm y}\hat{x} - k_{\rm x}\hat{y}$ is stable 
in accordance with the analytic solution of eq.~(\ref{eq:ap-Tc}), 
while the other pairing states are destabilized. 
 The anisotropy of the $d$-vector defined as $\Delta T_{\rm c}/T_{\rm c}$, 
is on the order of $O(1)$ for the realistic spin-orbit coupling 
$\bar{\alpha}^{2}/T_{\rm c0}t_{\rm z} \sim 10$. 
 This means that the $d$-vector is strongly pinned 
by the random spin-orbit coupling. 
 Note that this anisotropy is much larger than that 
in the clean centrosymmetric superconductors. 
 For example, we obtained a small anisotropy 
$\Delta T_{\rm c}/T_{\rm c} < 0.01$ for the clean bulk \Ruf, 
\cite{yanaseRu,udagawa2005} which has been confirmed 
experimentally. \cite{ishida1998,murakawa2007,murakawa2004}

 Here, we comment on the $d$-vector 
that is odd with respect to $\kz$, namely, $\d(\ktd,\kz) = - \d(\ktd,-\kz)$. 
 In this case, the intralayer Cooper pairing vanishes 
as $\d_{\rm 2d}(\ktd) =0$, and therefore, $\d_{\rm 3d}(\k) = \d(\k)$. 
 Then, stacking faults give rise to a strong pair-breaking effect 
through the first term of eq.~(\ref{eq:ap-Tc}), independent of the spin 
degree of freedom. (See the thin dashed line in Fig.~3.) 
 This means that the pairing states 
$\d = k_{\rm z} \hat{x}$ and $\d = k_{\rm z} \hat{z}$ 
proposed by Hasegawa and Taniguchi \cite{hasegawa2009} 
are unlikely to be realized in \Pt if stacking faults exist there.  
 We assume that the $d$-vector is even with respect to $\kz$ in the 
following part, unless we mention otherwise.

\subsection{Pair-breaking effect for $\d = k_{\rm y}\hat{x} - k_{\rm x}\hat{y}$}

 Next, we investigate the pair-breaking effect on the most stable 
pairing state $\d = k_{\rm y}\hat{x} - k_{\rm x}\hat{y}$. 
 When we assume a momentum dependence of the $d$-vector 
so that $\d(\k) \propto \g(\k)$ and $\d_{\rm 3d}(\k)=0$, as in \S4.1, 
the \Tc of the spin triplet pairing state 
$\d = k_{\rm y}\hat{x} - k_{\rm x}\hat{y}$ is not decreased 
by stacking faults. 
 On the other hand, a weak pair-breaking effect is induced for 
$\d = k_{\rm y}\hat{x} - k_{\rm x}\hat{y}$ 
through the first and second terms in eq.~(\ref{eq:ap-Tc}) 
when the momentum dependences of the $d$- and $g$-vectors 
are more complicated. 
 We discuss the following contributions; 
 (I) the first term of eq.~(\ref{eq:ap-Tc}), which arises from the 
interlayer Cooper pairing, and  
(II) the second term, which originates from the mismatch of the 
$d$- and $g$-vectors.

\begin{figure}[htbp]
  \begin{center}
\includegraphics[width=7cm]{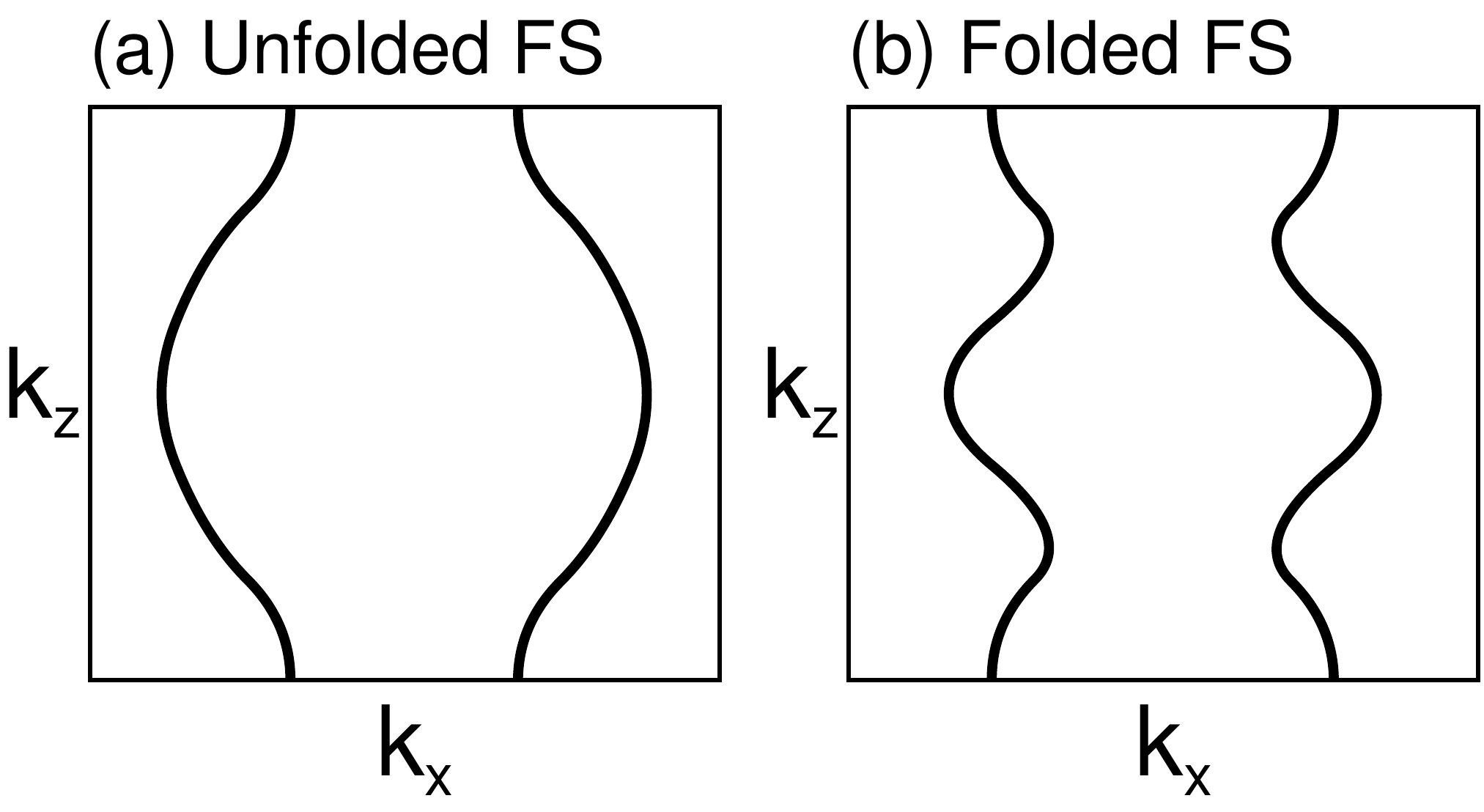}
\caption{
Examples of (a) unfolded and (b) folded Fermi surfaces along 
the $\kz$-axis. The cross-sections on the $k_{\rm x}$-$k_{\rm z}$ plane 
are shown. 
}
  \end{center}
\end{figure}

 (I) According to eq.~(\ref{eq:ap-Tc}), both random scalar potential and 
random spin-orbit coupling lead to the pair-breaking of interlayer 
Cooper pairs represented by $\d_{\rm 3d}(\k)$. 
 However, this pair-breaking effect vanishes for the simple band structure 
when the Fermi surface is not folded along the $\kz$-axis. 
 Examples of the unfolded and folded Fermi surfaces 
are shown in Figs.~4(a) and 4(b), respectively. 
 We explain this nontrivial result on the basis of the weak-coupling theory.
 Since \Tc is determined by the Cooper pairs on the Fermi surface 
in the weak-coupling limit, we replace eq.~(8) with, 
\begin{eqnarray}
\label{eq:ap-psi_3d}
&& \hspace*{-10mm}
\sum_{j=1,\pm}^{l} \d_{\rm 3d}(\ktd,\pm \kz^{j}) 
/|v_{\rm z}(\ktd,\pm \kz^{j})| = 0,  
\end{eqnarray}
where the momentum on the Fermi surface are described by $(\ktd,\pm \kz^{j})$. 
 For unfolded Fermi surfaces $l=1$ for all $\ktd$. 
 This is the case for eq.~(\ref{eq:dispersion}). 
 According to eq.~(\ref{eq:ap-psi_3d}), the three-dimensional component 
$\d_{\rm 3d}(\k)$ vanishes on the Fermi surface
when the order parameter is even with respect to $k_{\rm z}$ and  
$\d(\ktd,\kz) = \d(\ktd,-\kz)$. 
 Then, the pair-breaking effect through the first term 
in eq.~(\ref{eq:ap-Tc}) vanishes. 
 In other words, the spin triplet pairing state 
$\d = k_{\rm y}\hat{x} - k_{\rm x}\hat{y}$ is robust against 
stacking faults even for the three-dimensional order parameter 
and/or three-dimensional band structure when the Fermi surface is unfolded. 
 This is viewed as an extension of Anderson's theorem 
\cite{Anderson_impurity} for the non-$s$-wave superconductors.

 For a folded Fermi surface with $l \geq 2$ [Fig.~4(b)], 
pair-breaking occurs for $\d = k_{\rm y}\hat{x} - k_{\rm x}\hat{y}$ 
through the interlayer pairing $\d_{\rm 3d}(\k)$. 
 This effect is quantitatively important when the $d$-vector $\d(\k)$ 
changes its sign along the $\kz$-axis. In such a case, the 
horizontal line node (or a tiny gap) appears in the superconducting gap 
$\Delta(\k) \propto \sqrt{\phi_{\rm x}(\k)^{2}+\phi_{\rm y}(\k)^{2}}$ 
if the nodes of $\phi_{\rm x}(\k)$ and $\phi_{\rm y}(\k)$ are close to each other.

 The presence or absence of the pair-breaking effect due to the 
interlayer Cooper pairing is summarized in Table II. 
 The case of the odd $d$-vector with respect to $\kz$ is also shown 
in Table II. 
 We see that pair-breaking occurs only for the folded Fermi surface 
with horizontal line nodes of the superconducting gap when the 
order parameter is even with respect to $\kz$.

\begin{table}
\begin{tabular}[htb]{|c||c|c||c|} \hline
Parity for $\kz$ & \multicolumn{2}{|c||}{Even} & Odd \\ \hline
Gap structure & Full gap or Vertical & Horizontal & Horizontal 
\\ \hline \hline
Unfolded FS & $\times$ & $\times$  & $\bigcirc$ \\ \hline
Folded FS & $\times$ & $\bigcirc$  & $\bigcirc$ \\ \hline
\end{tabular}
\caption{Summary of the pair-breaking effect on the most stable 
spin triplet pairing state $\d = k_{\rm y}\hat{x} - k_{\rm x}\hat{y}$ 
through the interlayer Cooper pairing. 
 The second row indicates the gap structure. 
``Vertical'' and ``Horizontal'' denote the vertical and horizontal line nodes 
in the superconducting gap, respectively. 
 The third and fourth rows describe the folded and unfolded Fermi surfaces, 
respectively. 
 The symbols $\bigcirc$ and $\times$ show the presence and absence of 
the pair-breaking effect, respectively. 
 The fourth column shows the pair-breaking effect for the pairing state 
having odd parity with respect to $\kz$. 
}
\end{table}

 (II) The pair-breaking effect arises from the intralayer Cooper pairs 
when the momentum dependences of the $d$-vector and $g$-vector are mismatched. 
 Although short-range Cooper pairing leads to the simple 
momentum dependence of the $d$-vector, the $g$-vector may have 
a complicated momentum dependence. \cite{yanasencscfull} 
 Then, the $d$-vector cannot be parallel to the $g$-vector in the 
whole Brillouin zone. \cite{yanasencsclett,yanasencscfull,tada2009,yada2009}
 In such a case, \Tc is decreased even for the most stable pairing state 
$\d = k_{\rm y} \hat{x} - k_{\rm x} \hat{y}$ through the second term 
of eq.~(\ref{eq:ap-Tc}). 
 This is similar to the case of clean non-centrosymmetric superconductors, 
\cite{yanasencsclett,yanasencscfull} 
but the amplitude of the pair-breaking effect is reduced. 
 The pair-breaking effect due to the random spin-orbit 
coupling is quadratic in $\bar{\alpha}$, while it is linear in $\alpha$ 
in a clean non-centrosymmetric system. 
 In \S5.1, we show that this quantitative difference resolves the 
unusual variation of \Tc in \Ptf.

\subsection{Highly two-dimensional system}

 We here comment on the breakdown of the Born approximation in 
highly two-dimensional systems. 
 The parameter space for the {\it c}-axis kinetic energy $W_{\rm z}$ 
is divided into the following three regimes: 
 (A) Three-dimensional regime 
$\bar{\alpha}^{2}/E_{\rm F} < W_{\rm z}$ 
where the Born approximation is valid. 
 Because the phase relaxation rate is inversely proportional to 
$W_{\rm z}$, the pair-breaking effect increases with decreasing $W_{\rm z}$ 
by enhancing the two-dimensionality. 
 (B) Two-dimensional regime $T_{\rm c0} < W_{\rm z} < \bar{\alpha}^{2}/E_{\rm F}$
where the Born approximation breaks down. 
 The effects of the randomness tend to be saturated with 
decreasing $W_{\rm z}$.  
 A higher-order theory, such as the self-consistent 
T-matrix approximation, is needed for a quantitative estimation.  
 (C) Highly two-dimensional regime $W_{\rm z} < T_{\rm c0}$ 
where the spatial inhomogeneity plays an important role like in 
short coherence length superconductors 
\cite{franz1997,ghosal2000,yanase2006} 
 In this regime the spatial average (replica symmetry) taken in the 
Born approximation as well as in the self-consistent T-matrix 
approximation is not justified. 
 In the limit of two-dimensionality $W_{\rm z}/T_{\rm c0} \rightarrow 0$, 
the layers are independent of each other. 
 Then, the pair-breaking effect on \Tc vanishes, because \Tc 
is determined by the layer with $\alpha_{i}=0$. 
 Summarizing this discussion, we show the schematic figure in Fig.~5, 
where the pair-breaking effect is shown for regimes (A), (B), and (C).

\begin{figure}[htbp]
  \begin{center}
\includegraphics[width=6cm]{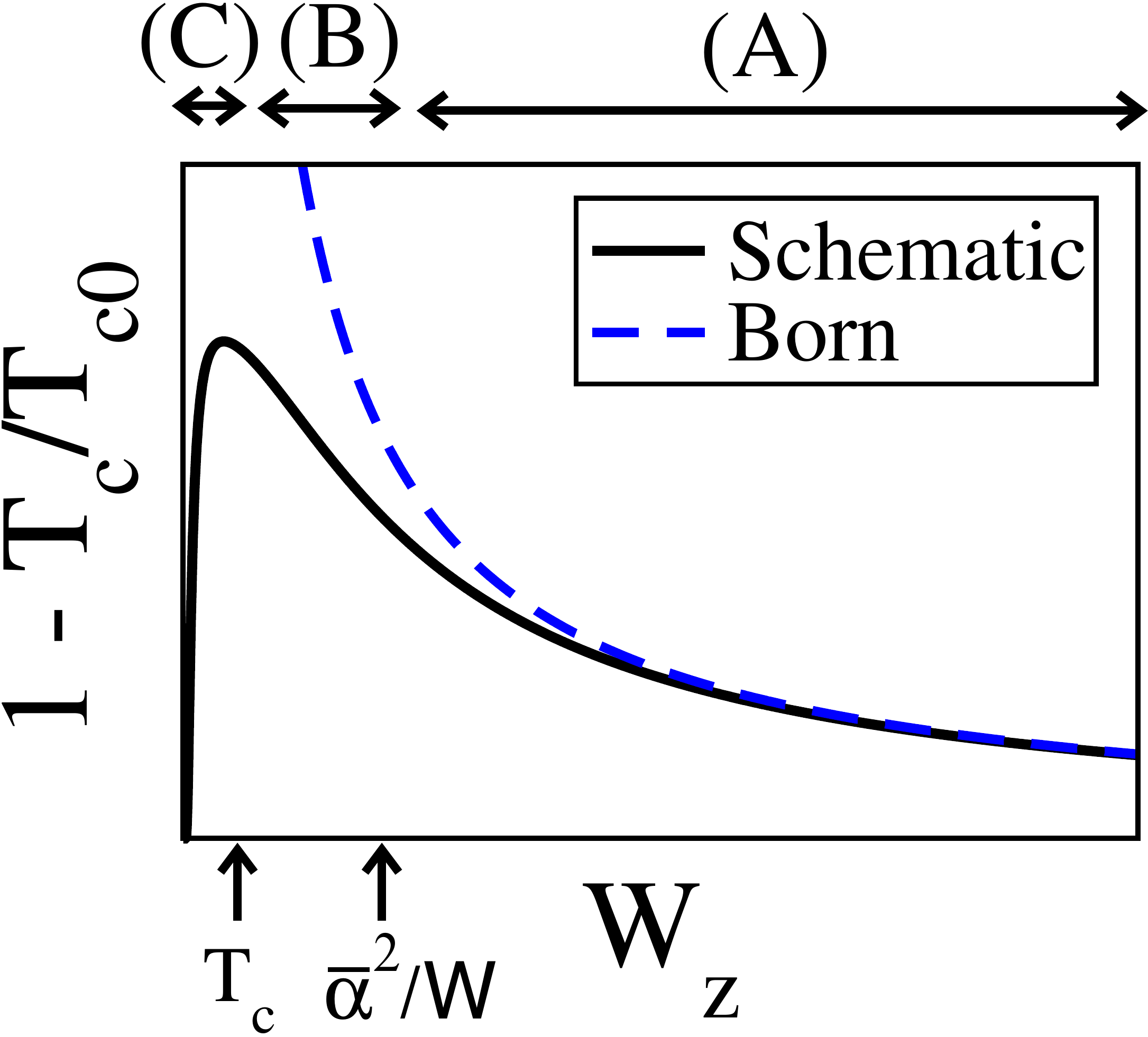}
\caption{(Color online) 
Schematic figure of the pair-breaking effect  
due to stacking faults (solid line). 
We show the three-dimensional regime (A), two-dimensional regime (B), 
and highly two-dimensional regime (C). 
The dashed line shows the result of the Born approximation, which is valid 
in regime (C). 
The details are explained in the text. 
}
  \end{center}
\end{figure}

\section{CePt$_3$Si and Sr$_2$RuO$_4$}

 We here turn to examples of possible spin triplet 
superconductors. \Pt and \Ru are discussed in \S5.1 and \S5.2, 
respectively.

\subsection{CePt$_3$Si}
 
 First, we show that an unresolved issue in \Pt is resolved by 
taking into account the randomness in the spin-orbit coupling.  
 After the discovery of superconductivity with $T_{\rm c} \sim 0.7$ K 
by Bauer {\it et al.}, \cite{bauer2004} another superconducting phase with 
$T_{\rm c} \sim 0.45$ K was found. \cite{takeuchi2007,motoyama2008} 
 Several experimental results show that the high-\Tc phase is more disordered 
than the low-\Tc phase. \cite{takeuchi2007,motoyama2008} 
 This variation of \Tc is unusual since the heavy fermion superconductor \Pt 
is considered to be a non-$s$-wave superconductor.  
 We here resolve this problem by assuming the presence of stacking faults 
in the high-\Tc phase, as proposed in ref.~30. 

 One of the important consequences of \S4 is the extended Anderson's 
theorem for stacking faults. 
 The non-$s$-wave superconductivity is robust against stacking faults 
in many cases, as summarized in Table II. 
 In particular, the pair-breaking effects due to the random scalar potential 
and the random spin-orbit coupling can be substantially avoided 
for the spin triplet pairing state with 
$\d = k_{\rm y} \hat{x} - k_{\rm x} \hat{y}$. 

 Another point is the weak but finite pair-breaking effect 
for $\d = k_{\rm y} \hat{x} - k_{\rm x} \hat{y}$ arising from both 
the uniform and random spin-orbit couplings. 
 Since our formulation does not include the uniform spin-orbit coupling, 
we cannot interpolate between the clean and random systems. 
 However, we can compare \Tc in the clean limit with that in the 
highly disordered system since \Tc in the clean non-centrosymmetric 
superconductor is obtained by replacing 
$\Gamma^{\alpha}(\ktd) = \pi \bar{\alpha}^{2} |\g(\ktd)|^{2} \rho^{\rm z} (\ktd)$ 
in eq.~(\ref{eq:ap-Tc}) 
with $\Gamma^{\alpha}(\ktd) = \alpha |\g(\ktd)|$. \cite{frigeriprl} 
 Because the relations 
$\bar{\alpha}/W_{\rm z} \ll 1$ and $\bar{\alpha} \leq \alpha$ 
are satisfied in \Ptf, the pair-breaking effect is larger in the 
clean \Pt than in the disordered \Ptf. 
 In other words, \Tc is increased by stacking faults 
by recovering the global inversion symmetry. 
 This is consistent with the seemingly unusual variation of \Tc 
in \Ptf. \cite{takeuchi2007,motoyama2008}

 In order to examine this proposal quantitatively, 
we take into account the band structure of \Pt and  
numerically solve eqs.~(\ref{eq:Green-function})-(\ref{eq:Tc}) for 
the following dispersion relation: 
\begin{eqnarray}
\label{eq:dispersion_CePt3Si}
&& \hspace*{-11mm}  \e(\k)  =   2 t_1 (\cos k_{\rm x} +\cos k_{\rm y}) 
         + 4 t_2 \cos k_{\rm x} \cos k_{\rm y} 
\nonumber \\ && \hspace*{-6mm} 
         + 2 t_3 (\cos 2 k_{\rm x} +\cos 2 k_{\rm y})
+ [ 2 t_4 + 4 t_5 (\cos k_{\rm x} +\cos k_{\rm y}) 
\nonumber \\ && \hspace*{-6mm} 
         + 4 t_6 (\cos 2 k_{\rm x} +\cos 2 k_{\rm y}) ] \cos k_{\rm z}
         + 2 t_7 \cos 2 k_{\rm z} 
         - \mu. 
\end{eqnarray}
 By choosing the parameters as 
$(t_1,t_2,t_3,t_4,t_5,t_6,t_7,n) = (1,-0.15,-0.5,-0.3,-0.1,-0.09,-0.2,1.75)$, 
eq.~(\ref{eq:dispersion_CePt3Si}) reproduces the $\beta$-band of CePt$_3$Si.  
\cite{yanasencsclett} 
 Although \Pt has several Fermi surfaces, 
\cite{samokhin2004,anisimovprivate,hashimoto2004} 
it is expected that the superconductivity is mainly induced 
by the $\beta$-band since the $\beta$-band has a substantial 
Ce 4$f$-electron character~\cite{anisimovprivate} and 
the largest density of states. \cite{samokhin2004} 
 The Fermi surface obtained from eq.~(\ref{eq:dispersion_CePt3Si}) 
(see Fig.~1 of ref.~16) is folded 
along the $\kz$-axis for a part of $\ktd$. 
 Therefore, not only the random spin-orbit coupling but also the 
random scalar potential can decrease \Tcf.

 We assume the $g$-vector 
\begin{eqnarray}
\label{eq:g-vector-CePt3Si}
&& \hspace*{-8mm}
\g(\k) = 
(-\sin k_{\rm y} (1 - G \sin k_{\rm x}^{2}), 
\sin k_{\rm x} (1 - G \sin k_{\rm y}^{2}), 0)
\nonumber \\ && \hspace*{3mm}
/<|\g(\k)|>_{\rm F}, 
\end{eqnarray}
%
while the $d$-vector is assumed to be $\d(\k) = 
\frac{1}{\sqrt{2}} (\phi_{\rm y}(\k),- \phi_{\rm x}(\k),0)$ with 
\begin{eqnarray}
\label{eq:d-vector-CePt3Si}
&& \hspace*{-15mm}
(\phi_{\rm x}(\k),\phi_{\rm y}(\k)) = 
(1 + D \cos k_{\rm z}) \times (\sin k_{\rm x},\sin k_{\rm y}). 
\end{eqnarray}
The parameter $G$ represents the complexity of the $g$-vector, while 
the parameter $D$ represents the weight of interlayer Cooper pairing.

\begin{figure}[htbp]
  \begin{center}
\includegraphics[width=7cm]{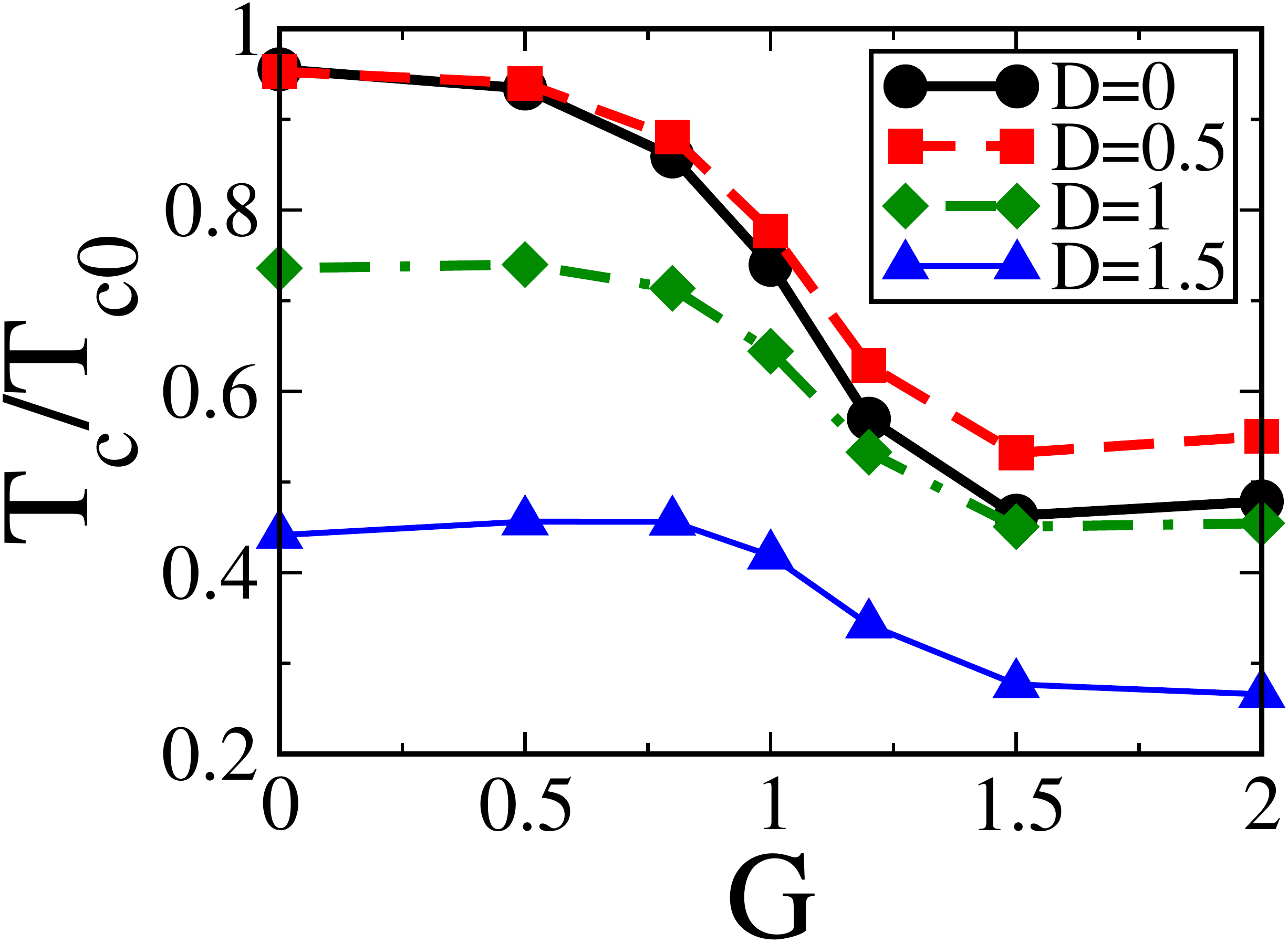}
\caption{(Color online) 
$T_{\rm c}$ values of the spin triplet pairing state 
$\d = k_{\rm y} \hat{x} - k_{\rm x} \hat{y}$
for various values of $G$ and $D$. 
The circles, squares, diamonds, and triangles show 
the $G$ dependences for $D=0$, $0.5$, $1$, and $1.5$, respectively. 
The dispersion relation in eq.~(\ref{eq:dispersion_CePt3Si}) is assumed. 
We fix the randomness $\Gamma^{u}/T_{\rm c0}=\Gamma^{\alpha}/T_{\rm c0}=5$ with 
$\Gamma^{u}=\bar{u}^{2}/|t_{4}|$ and $\Gamma^{\alpha} = \bar{\alpha}^{2}/|t_{4}|$. 
\Tc in the clean limit without spin-orbit coupling is assumed to be 
$T_{\rm c0} = 0.0064$. 
}
  \end{center}
\end{figure}

 We show the variation of \Tc with respect to $D$ and $G$ in Fig.~6, 
where the randomness is fixed to be 
$\bar{u}^{2}/T_{\rm c0}|t_{4}| = \bar{\alpha}^{2}/T_{\rm c0}|t_{4}| = 5$. 
 It is shown that \Tc is significantly decreased with increasing $G$ 
at approximately $G=1$. 
 This is because nontrivial topological defects appear in 
the $g$-vector for $G > 1$. 
 Thus, the \Tc of spin triplet superconductivity 
with $\d = k_{\rm y} \hat{x} - k_{\rm x} \hat{y}$ is substantially decreased 
by the random spin-orbit coupling when the topological properties are 
different between the $d$-vector and $g$-vector. 
 Another effect of the topological defects in the $g$-vector, 
such as the topologically protected line node of the superconducting gap, 
has been pointed out. \cite{yanasencsclett,yanasencscfull} 
 For the dependence on $D$, we see a substantial decrease in \Tc 
for $D \geq 1$. This is because the superconducting gap has horizontal 
line nodes for $D \geq 1$, and then the superconductivity is 
suppressed by stacking faults in accordance with Table II.

\begin{figure}[htbp]
  \begin{center}
\includegraphics[width=7cm]{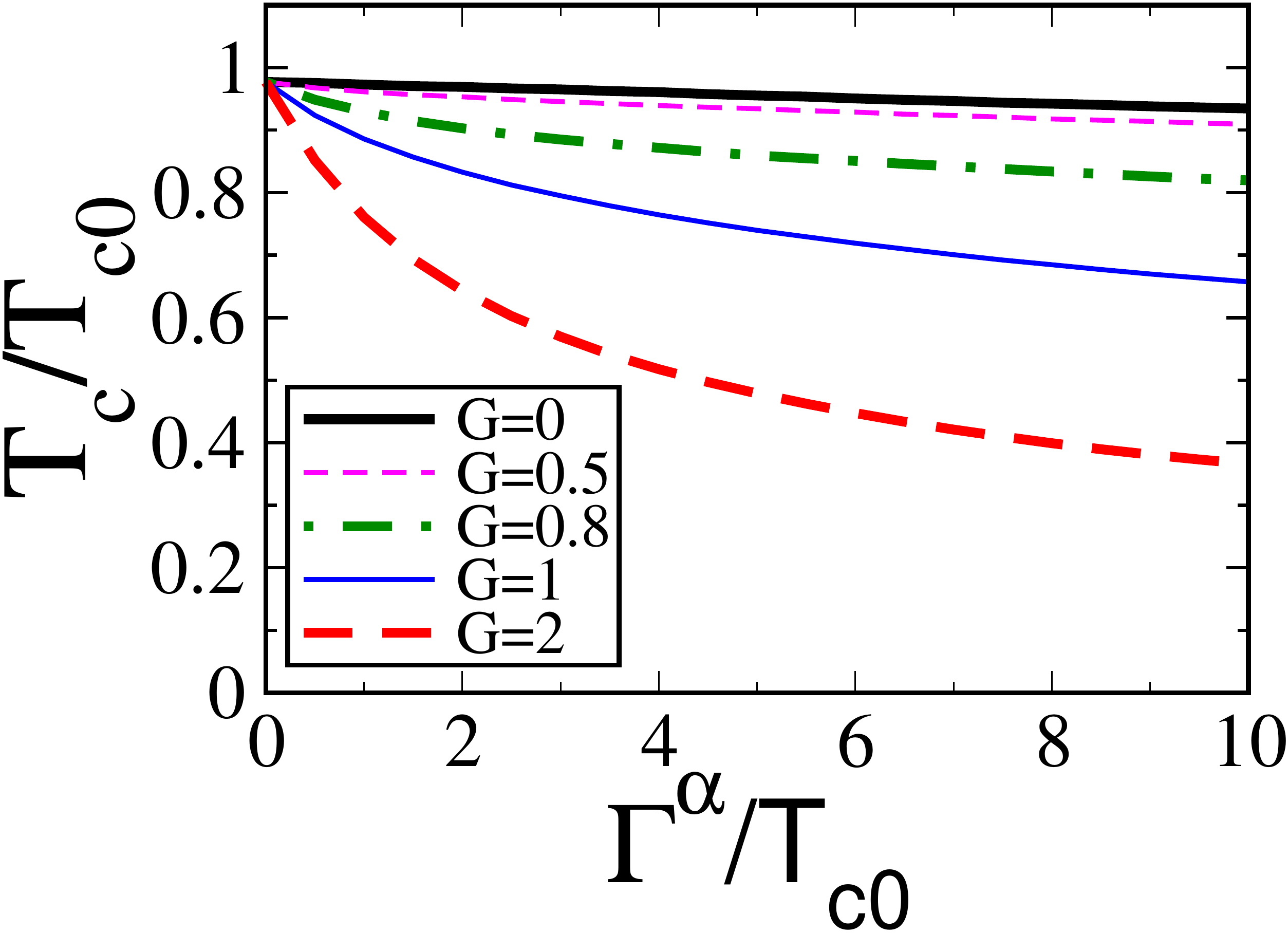}
\caption{(Color online) 
The $T_{\rm c}$ of the spin triplet pairing state 
$\d = k_{\rm y} \hat{x} - k_{\rm x} \hat{y}$ for various values of $G$ and 
$\Gamma^{\alpha}$. 
 The thick solid, thin dashed, dash-dotted, thin solid, and 
thick dashed lines show the $\Gamma^{\alpha}$-dependence of \Tc 
for $G=0$, $0.5$, $0.8$, $1$, and $2$, respectively. 
We here assume $D=0$.  
The other parameters are the same as in Fig.~6. 
}
  \end{center}
\end{figure}

 For a quantitative comparison with experiments, 
we discuss the realistic parameters for \Ptf. 
 According to the microscopic analysis based on the random phase 
approximation, the interlayer Cooper pairing is negligible 
in the $s$+$P$-wave state of \Ptf. \cite{yanasencsclett,yanasencscfull} 
 This is the dominantly spin triplet pairing state 
consistent with the experimental results. \cite{sigristncsc} 
 Thus, the small parameter $D \ll 1$ is indicated in \Ptf. 
 On the other hand, it is difficult to determine the parameter $G$ 
since the momentum dependence of the $g$-vector has not been extracted from 
the data of band calculation. \cite{samokhin2004,anisimovprivate,hashimoto2004} 
 Therefore, we assume $D=0$ and show the $\bar{\alpha}$ dependence of \Tc 
for various $G$ in Fig.~7. 

 We here assume $\alpha/T_{\rm c0} = 100$ and $\alpha/|t_{4}| = 0.2$, 
where $\alpha$ is the spin-orbit coupling in the clean limit and 
the $T_{\rm c0}$ is the fictitious transition temperature for $\alpha=0$. 
 For the disordered phase of \Ptf, we assume $\bar{\alpha} = \alpha/2$,  
and then we obtain $\Gamma^{\alpha}/T_{\rm c0} = 5$.  
 \Tc in the clean limit of \Pt is roughly estimated by 
replacing $\Gamma^{\alpha}/T_{\rm c0}$ in Fig.~7 with $\alpha/T_{\rm c0}$. 
 When we assume a moderately complicated $g$-vector with $G=1$, 
we obtain $T_{\rm c}/T_{\rm c0} = 0.74$ for the disordered phase and 
$T_{\rm c}/T_{\rm c0} = 0.32$ in the clean limit. 
 This rough estimation is in reasonable agreement with 
the high \Tc of the disordered \Pt ($T_{\rm c} =0.7$ K) and 
the low \Tc of the clean \Pt ($T_{\rm c} =0.45$ K). 
 Thus, the seemingly unusual variation of \Tc in \Pt is understood 
by taking into account the spin-orbit coupling 
and assuming the $d$-vector $\d = k_{\rm y} \hat{x} - k_{\rm x} \hat{y}$. 
 We stress again that \Tc is increased in the disordered phase 
by recovering the global inversion symmetry.

\subsection{Sr$_2$RuO$_4$}

 Next, we discuss the superconductivity in  
bulk \Ru and the eutectic crystal \Ruf-\Rumf. 
 The superconductivity in bulk \Ru with $T_{\rm c}=1.5$ K 
was discovered in 1994. \cite{maeno1994}
%
 Recently, a new superconducting material has been fabricated in the 
eutectic crystal \Ruf-\Rumf. \cite{kittaka2008,kittaka2009,fittipaldi2008} 
 It has been indicated that the superconductivity occurs in the thin 
\Ru layers included in the \Rum region. \cite{kittaka2009,fittipaldi2008} 
 The disorder in the two-dimensional layer is expected to be negligible 
because the \Tc of \Ruf-\Rum is similar to that of bulk \Ruf. 
 Therefore, \Ruf-\Rum can be regarded as a layered ruthenate 
with many stacking faults, as shown in Fig.~1(b).

 For the bulk \Ruf, the $d$-vector has been theoretically investigated 
on the basis of the multi-orbital Hubbard model \cite{yanaseRu,nomuraprivate} 
and multi-orbital $d$-$p$ model. \cite{yoshioka2009,nomuraprivate} 
 Using these microscopic theories it is found that the anisotropy of 
the $d$-vector is very small, $\Delta T_{\rm c}/T_{\rm c} < 0.01$. 
 This is because the effect of the spin-orbit coupling ($L$-$S$ coupling) 
$\lambda$ on the superconductivity in the active $\gamma$-band is 
on the order of $\lambda^{2}/E_{\rm F}^{2}$. \cite{yanaseRu} 
 Such a small anisotropy is consistent with the NMR measurements 
\cite{murakawa2007,murakawa2004} 
and results in multiple phase transitions 
in the magnetic field. \cite{deguchi2002,udagawa2005}

 According to the results in \S4, the structure of the $d$-vector 
in \Ruf-\Rum is considerably different from that in the bulk \Ruf. 
 Several pieces of experimental evidence have been obtained for the 
chiral spin triplet pairing state 
$\d = (k_{\rm x} \pm {\rm i} k_{\rm y}) \hat{z}$ 
in the bulk \Ruf. \cite{Mackenzie2003} 
 On the other hand, the stable pairing state is expected to be 
$\d = k_{\rm y} \hat{x} - k_{\rm x} \hat{y}$ in \Ruf-\Rum 
owing to the random spin-orbit coupling arising from stacking faults.

 A weak anisotropy due to the $L$-$S$ coupling $\Delta T_{\rm c}/T_{\rm c} < 0.01$ 
is compensated for by the very small random antisymmetric spin-orbit coupling 
$\bar{\alpha}$ with $\bar{\alpha}^{2}/W_{\rm z}T_{\rm c0} < 0.02$  
according to Fig.~3. 
 When we assume $W_{\rm z} = 100$ K and $T_{\rm c0}=1.5$ K, the pairing state 
$\d = k_{\rm y} \hat{x} - k_{\rm x} \hat{y}$ 
is more stable than the chiral state for $\bar{\alpha} > 2$ K. 
 Since $\bar{\alpha} = 2$ K is much smaller than the typical 
antisymmetric spin-orbit coupling ($> 100$ K), 
the $d$-vector $\d = k_{\rm y} \hat{x} - k_{\rm x} \hat{y}$ 
is likely to be realized in \Ruf-\Rumf. 
 The anisotropy of the $d$-vector is expected to be large on the order of 
$\Delta T_{\rm c}/T_{\rm c} =O(1)$, 
since the relation $\bar{\alpha}^{2}/W_{\rm z}T_{\rm c0} > 1$ is expected. 
 This means that the spin triplet pairing state 
$\d = k_{\rm y} \hat{x} - k_{\rm x} \hat{y}$ with conserved 
time-reversal symmetry is very robust in the eutectic crystal 
\Ruf-\Rumf.

 The Born approximation may break down in \Ruf-\Rumf, as discussed in 
\S4.3, because the \Ru layers are dilute in the superconducting region 
of interest. 
 However, these results for the $d$-vector are qualitatively valid 
beyond the Born approximation.

 We here give a brief comment on the $3$ K superconducting phase of \Ruf,   
which is realized near the interface with the Ru metal.~\cite{maeno1998} 
 The antisymmetric spin-orbit coupling should play an important role 
in such inhomogeneous spin triplet superconductors because 
the local inversion symmetry is broken. 
 In particular, the spatial dependence of the $d$-vector is determined 
by the shape of the Ru metal. 
 Then, novel topological defects should appear for some structures 
of interfaces. 
 We leave such an interesting texture of a 
spin triplet order parameter as a future issue.

\section{Summary of $D$-vector in Spin Triplet Superconductors}

 Combining this study on disordered superconductors 
with the previous studies on clean bulk superconductors, 
we summarize the structures of the $d$-vector in spin triplet 
superconductors. 

 The $d$-vector in the clean centrosymmetric superconductors is determined 
by the symmetries of the crystal structure, local electron orbital, 
and superconductivity in accordance with the selection rules 
summarized in Table I. \cite{yanaseRu,yanaseCo} 
 We stress again that these results are exact in the lowest order of 
$\lambda/E_{\rm F}$. We also mentioned the cases where the higher-order 
terms with respect to $\lambda/E_{\rm F}$ determine the $d$-vector.  
 It is expected that similar selection rules will also be obtained 
for the $f$-electron systems in which the other limit 
$\lambda/E_{\rm F} \gg 1$ is appropriate, although the microscopic study 
of heavy fermions remains a future work.

\begin{table}
\begin{tabular}[htb]{|c|c|} \hline
Non-centrosymmetric  
& Directional disorder 
\\ \hline  \hline 
\multicolumn{2}{|c|}{$ \d \parallel \g$} 
\\ \hline 
min[$O(1)$,$O(\alpha/T_{\rm c0})$] 
& min[$O(1)$,$O(\bar{\alpha}^{2}/W T_{\rm c0})$] \\ \hline
\end{tabular}
\caption{Summary of $d$-vector in the non-centrosymmetric superconductors 
(first column) and disordered superconductors (second column). 
The $d$-vector (second row) is determined solely by the crystal structure 
that gives rise to the uniform and random antisymmetric spin-orbit coupling. 
The third row shows the anisotropy of the $d$-vector. 
See the text for details. 
}
\end{table}

 The $d$-vector in the clean non-centrosymmetric superconductors and 
disordered superconductors are summarized in Table III. 
 In these cases, the $d$-vector is determined solely by the crystal 
structure through the $g$-vector of antisymmetric spin-orbit coupling. 
 The anisotropy of the $d$-vector $\Delta T_{\rm c}/T_{\rm c}$ is estimated 
to be min[$O(1)$,$O(\alpha/T_{\rm c0})$] and  
min[$O(1)$,$O(\bar{\alpha}^{2}/E_{\rm F} T_{\rm c0})$] 
for the non-centrosymmetric superconductors and 
disordered superconductors, respectively. 
 Although the role of the antisymmetric spin-orbit coupling is reduced 
by the randomness, it is still much larger than the effect of the symmetric 
$L$-$S$ coupling.  
 This means that we have to be careful in discussing the $d$-vector of 
centrosymmetric spin triplet superconductors because 
it is affected by a small amount of directional disorders. 
 Tables I and III show a complete set of theoretical results on the 
structures of the $d$-vector.

\section{Summary and Discussion}

 We investigated the roles of random spin-orbit coupling in spin triplet 
superconductors. 
 The random antisymmetric spin-orbit coupling induced by stacking faults 
in \Pt and \Ruf-\Rum has been studied as a typical example. 
 It is shown that the $d$-vector parallel to the $g$-vector is stabilized by 
the spin-orbit coupling similarly to that in the non-centrosymmetric 
superconductors. 
 In the cases of \Pt and \Ruf-\Rumf, the pairing state  
$\d = k_{\rm y} \hat{x} - k_{\rm x} \hat{y}$ is stabilized. 
 The anisotropy of the $d$-vector 
is represented by the parameter $\bar{\alpha}^{2}/W_{\rm z}T_{\rm c0}$, 
which is much smaller than that in the clean non-centrosymmetric superconductors  
by the factor $\bar{\alpha}/W_{\rm z}$, but much larger than that in the clean 
centrosymmetric superconductors.

 The superconducting state of \Pt and \Ruf-\Rum was  
discussed on the basis of the stacking fault model. 
 A seemingly controversial issue of \Ptf, namely, the high \Tc of 
the disordered phase, has been resolved. 
 This unusual variation of \Tc is attributed to the restoration of 
the global inversion symmetry by disorders 
while keeping the broken local inversion symmetry.

 Stacking faults have more interesting effects on \Ruf-\Rumf. 
 While the chiral state $\d = (k_{\rm x} \pm {\rm i} k_{\rm y}) \hat{z}$ with 
broken time-reversal symmetry is considered to be realized in the bulk \Ruf, 
the pairing state $\d = k_{\rm y} \hat{x} - k_{\rm x} \hat{y}$ with 
time-reversal symmetry is likely to be stabilized in \Ruf-\Rumf. 
 Therefore, several properties of the superconducting state are 
considerably different between the eutectic crystal \Ruf-\Rum and 
the bulk \Ruf. 
 The comparison of these related materials provides an opportunity to 
study the multicomponent order parameters of \Ruf. 
 For example, the multiple phase transitions observed in 
the bulk \Ruf \cite{deguchi2002,tenya2006} should disappear in \Ruf-\Rumf.

 Since the $g$-vector of antisymmetric spin-orbit coupling is a real vector, 
the time-reversal symmetry is generally conserved 
in the spin triplet superconductors with directional disorders. 
 This is a means to realize topological superconductors 
having the time-reversal symmetry, whose nontrivial properties such as 
Majorana fermions and non-Abelian statistics are attracting growing attention. 
\cite{schnyder2008,roy2008,qi2009,sato2009}

 The directional disorders generally play an important role in the 
spin triplet superconductors, as discussed in this paper. 
 In particular, the random spin-orbit coupling due to the local 
inversion symmetry breaking can alter the pairing state 
in the centrosymmetric systems. 
 This is the first study pointing out that not only the broken {\it global} 
inversion symmetry but also the broken {\it local} inversion symmetry 
plays essential roles. 
 Although we focused on a particular example, that is, stacking faults, 
it is straightforward to extend this study to other cases such as the 
bilayer structure and pyrochlore structure. 
 Our study indicates that it is not difficult to study the $d$-vector 
from the theoretical point of view,  
because it is determined solely by the crystal structure in many cases.

\section*{Acknowledgements}
 The authors are grateful to S. Fujimoto, 
Y. Kitaoka, S. Kittaka, G. Motoyama, Y. Maeno, H. Mukuda, T. Nomura, 
Y. Onuki, M. Sigrist, R. Shindou, H. Yaguchi, and M. Yogi 
for fruitful discussions. 
 This work was supported by a 
Grant-in-Aid for Scientific Research on Priority Area 
``Superclean Materials'' (No. 20029008)
and a Grant-in-Aid for Scientific Research on Innovative Areas
``Heavy Electrons'' (No. 21102506) 
from MEXT, Japan. 
It was also supported by a Grant-in-Aid for 
Young Scientists (B) (No. 20740187) from JSPS. 
 Numerical computation in this work was carried out 
at the Yukawa Institute Computer Facility.

\bibliographystyle{jpsj}
\bibliography{64337}

\begin{thebibliography}{10}

\bibitem{Leggett1975}
A.~J. Leggett:  Rev. Mod. Phys. {\bf 47} (1975) 331.
\bibitem{fisher1989}
R.~A. Fisher, S. Kim, B.~F. Woodfield, N.~E. Phillips, L. Taillefer, K.
  Hasselbach, J. Flouquet, A.~L. Giorgi, and J.~L. Smith:  Phys. Rev. Lett.
  {\bf 62} (1989) 1411.
\bibitem{Sigrist1991}
M. Sigrist and K. Ueda:  Rev. Mod. Phys. {\bf 63} (1991) 239.
\bibitem{Tou1998}
H. Tou, Y. Kitaoka, K. Ishida, K. Asayama, N. Kimura, Y. Onuki, E. Yamamoto, Y.
  Haga, and K. Maezawa:  Phys. Rev. Lett. {\bf 80} (1998) 3129.
\bibitem{Tou1996}
H. Tou, Y. Kitaoka, K. Asayama, N. Kimura, Y. Onuki, E. Yamamoto, and K.
  Maezawa:  Phys. Rev. Lett. {\bf 77} (1996) 1374.
\bibitem{Machida1999}
K. Machida, T. Nishira, and T. Ohmi:  J. Phys. Soc. Jpn. 
  {\bf 68} (1999) 3364.
\bibitem{Sauls1994}
J.~A. Sauls:  Adv. Phys. {\bf 43} (1994) 113.
\bibitem{Joynt2002}
R. Joynt and L. Taillefer:  Rev. Mod. Phys. {\bf 74} (2002) 235.
\bibitem{maeno1994}
Y. Maeno, H. Hashimoto, K. Yoshida, S. Nishizaki, T. Fujita, J.~G. Bednorz, and
F. Lichtenberg:  Nature {\bf 372} (1994) 532.
\bibitem{bauer2004}
E. Bauer, G. Hilscher, H. Michor, C. Paul, E.~W. Scheidt, A. Gribanov, Y.
  Seropegin, H. No\"el, M. Sigrist, and P. Rogl:  Phys. Rev. Lett. {\bf 92}
  (2004) 027003.
\bibitem{bauerreview}
E. Bauer, H. Kaldarar, A. Prokofiev, E. Royanian, A. Amato, J. Sereni, W.
  Br\"{a}mer-Escamilla, and I. Bonalde: 
J. Phys. Soc. Jpn. {\bf 76} (2007) 051009.
\bibitem{Mackenzie2003}
A.~P. Mackenzie and Y. Maeno:  Rev. Mod. Phys. {\bf 75} (2003) 657.
\bibitem{SigristRu}
M. Sigrist:  Prog. Theor. Phys. Suppl. {\bf 160} (2005) 1.
\bibitem{sigristncsc}
M. Sigrist, D.~F. Agterberg, P.~A. Frigeri, N. Hayashi, R.~P. Kaur, A. Koga, I.
  Milat, K. Wakabayashi, and Y. Yanase: 
{J. Magn. Magn. Mater.} {\bf {310}} ({2007}) {536}.
\bibitem{hayashi2006}
N. Hayashi, K. Wakabayashi, P. A. Frigeri, and M. Sigrist: 
Phys. Rev. B {\bf 73} (2006) 024504. 
\bibitem{yanasencsclett}
Y. Yanase and M. Sigrist:  J. Phys. Soc. Jpn.  {\bf 76}
  (2007) 043712.
\bibitem{yanasencschelical}
Y. Yanase and M. Sigrist:  J. Phys. Soc. Jpn.  {\bf 76}
  (2007) 124709.
\bibitem{yanasencscfull}
Y. Yanase and M. Sigrist:  J. Phys. Soc. Jpn.  {\bf 77}
  (2008) 124711.
\bibitem{frigeriprl}
P.~A. Frigeri, D.~F. Agterberg, A. Koga, and M. Sigrist:  Phys. Rev. Lett. {\bf
  92} (2004) 097001.
\bibitem{fujimotoreview}
S. Fujimoto:  J. Phys. Soc. Jpn.  {\bf 76} (2007) 051008.
\bibitem{edelstein1989}
V.~M. E\'del\'shtein:  Sov. Phys. JETP {\bf 68} (1989) 1244.
\bibitem{gorkov-rashba}
L.~P. Gor'kov and E.~I. Rashba:  Phys. Rev. Lett. {\bf 87} (2001) 037004.
\bibitem{yanaseRu}
Y. Yanase and M. Ogata:  J. Phys. Soc. Jpn.  {\bf 72}
  (2003) 673.
\bibitem{yanaseCo}
Y. Yanase, M. Mochizuki, and M. Ogata:  J. Phys. Soc. Jpn. 
{\bf 74} (2005) 2568.
\bibitem{murakawa2007}
H. Murakawa, K. Ishida, K. Kitagawa, H. Ikeda, Z.~Q. Mao, and Y. Maeno:
  J. Phys. Soc. Jpn.  {\bf 76} (2007) 024716.
\bibitem{murakawa2004}
H. Murakawa, K. Ishida, K. Kitagawa, Z.~Q. Mao, and Y. Maeno:  Phys. Rev. Lett.
  {\bf 93} (2004) 167004.
\bibitem{kittaka2008}
S. Kittaka, S. Fusanobori, S. Yonezawa, H. Yaguchi, Y. Maeno, R. Fittipaldi,
  and A. Vecchione:  Phys. Rev. B
  {\bf 77} (2008) 214511. 
\bibitem{kittaka2009}
S. Kittaka, S. Yonezawa, H. Yaguchi, Y. Maeno, R. Fittipaldi, A. Vecchione,
  J.~F. Mercure, A. Gibbs, R.~S. Perry, and A.~P. Mackenzie:  
J. Phys.: Conf. Ser. {\bf 150} (2009) 052113.
\bibitem{fittipaldi2008}
R. Fittipaldi, A. Vecchione, R. Ciancio, S. Pace, M. Cuoco, D. Stornaiuolo, D.
  Born, F. Tafuri, E. Olsson, S. Kittaka, H. Yaguchi, and Y. Maeno: 
Europhys. Lett. {\bf 83} (2008) 27007.
\bibitem{mukuda2009}
H. Mukuda, S. Nishide, A. Harada, K. Iwasaki, M. Yogi, M. Yashima, Y. Kitaoka,
  M. Tsujino, T. Takeuchi, R. Settai, Y. \={O}nuki, E. Bauer, K.~M. Itoh, and
  E.~E. Haller:  J. Phys. Soc. Jpn.  {\bf 78} (2009)
  014705.
\bibitem{takeuchi2007}
T. Takeuchi, T. Yasuda, M. Tsujino, H. Shishido, R. Settai, H. Harima, and Y.
  \={O}nuki:  J. Phys. Soc. Jpn.  {\bf 76} (2007) 014702.
\bibitem{motoyama2008}
G. Motoyama, K. Maeda, and Y. Oda:  J. Phys. Soc. Jpn. 
  {\bf 77} (2008) 044710.
\bibitem{read-green}
N. Read and D. Green:  Phys. Rev. B {\bf 61} (2000) 10267.
\bibitem{ivanov2001}
D.~A. Ivanov:  Phys. Rev. Lett. {\bf 86} (2001) 268.
\bibitem{schnyder2008}
A.~P. Schnyder, S. Ryu, A. Furusaki, and A.~W.~W. Ludwig:  Phys. Rev. B 
{\bf 78} (2008) 195125.
\bibitem{roy2008}
R. Roy:  arXiv:0803.2868.
\bibitem{qi2009}
X.-L. Qi, T.~L. Hughes, S. Raghu, and S.-C. Zhang:  Phys. Rev. Lett. {\bf 102}
  (2009) 187001.
\bibitem{sato2009}
M. Sato:  Phys. Rev. B {\bf 79} (2009) 214526.
\bibitem{nomuraprivate}
T. Nomura and H. Ikeda: private communication.
\bibitem{yoshioka2009}
Y. Yoshioka and K. Miyake:  J. Phys. Soc. Jpn.  {\bf 78}
  (2009) 074701.
\bibitem{rashba1959}
E.~I. Rashba:  Sov. Phys. Solid State {\bf 1} (1959) 368.
\bibitem{udagawa2005}
M. Udagawa, Y. Yanase, and M. Ogata:  J. Phys. Soc. Jpn. 
  {\bf 74} (2005) 2905.
\bibitem{ishida1998}
K. Ishida, H. Mukuda, Y. Kitaoka, K. Asayama, Z.~Q. Mao, Y. Mori, and Y. Maeno:
   Nature {\bf 396} (1998) 658.
\bibitem{hasegawa2009}
Y. Hasegawa and H. Taniguchi:  J. Phys. Soc. Jpn.  {\bf 78} (2009) 074717.
\bibitem{Anderson_impurity}
P. W. Anderson:  J. Phys. Chem. Solids {\bf 11} (1959) 26.
\bibitem{tada2009}
Y. Tada, N. Kawakami, and S. Fujimoto:  
New J. Phys. {\bf 11} (2009) 055070.
\bibitem{yada2009}
K. Yada, S. Onari, Y. Tanaka, and J. Inoue:  Phys. Rev. B {\bf 80} (2009)
  140509.
\bibitem{franz1997}
M. Franz, C. Kallin, A.~J. Berlinsky, and M.~I. Salkola:  Phys. Rev. B {\bf 56}
  (1997) 7882.
\bibitem{ghosal2000}
A. Ghosal, M. Randeria, and N. Trivedi:  Phys. Rev. B {\bf 63} (2000) 020505.
\bibitem{yanase2006}
Y. Yanase:  J. Phys. Soc. Jpn.  {\bf 75} (2006) 124715.
\bibitem{samokhin2004}
K.~V. Samokhin, E.~S. Zijlstra, and S.~K. Bose:  Phys. Rev. B {\bf 69} (2004)
  094514.
\bibitem{anisimovprivate}
A. Kozhevnikov and V. Anisimov: private communication.
\bibitem{hashimoto2004}
S. Hashimoto, T. Yasuda, T. Kubo, H. Shishido, T. Ueda, R. Settai, T.~D.
  Matsuda, Y. Haga, H. Harima, and Y. Onuki:  J. Phys.: Condens. Matter 
{\bf 16} (2004) L287. 
\bibitem{deguchi2002}
K. Deguchi, M.~A. Tanatar, Z. Mao, T. Ishiguro, and Y. Maeno:  
J. Phys. Soc. Jpn. {\bf 71} (2002) 2839.
\bibitem{maeno1998}
Y. Maeno, T. Ando, Y. Mori, E. Ohmichi, S. Ikeda, S. NishiZaki, 
and S. Nakatsuji: Phys. Rev. Lett. {\bf 81} (1998) 3765. 
\bibitem{tenya2006}
K. Tenya, S. Yasuda, M. Yokoyama, H. Amitsuka, K. Deguchi, and Y. Maeno:
  J. Phys. Soc. Jpn.  {\bf 75} (2006) 023702.
\end{thebibliography}

\end{document}